\documentclass[12pt,preprint]{aastex}
\slugcomment{Draft Version Jun 22, 2007}

\shorttitle{Triggering Mechanism for a Filament Eruption} 

\begin{document}

\title{Triggering Mechanism for the Filament Eruption 
on 2005 September 13
in Active Region NOAA 10808}

\author{Kaori Nagashima\altaffilmark{1,2}, 
Hiroaki Isobe\altaffilmark{3},
Takaaki Yokoyama\altaffilmark{3},
Takako T. Ishii\altaffilmark{2},
Takenori J. Okamoto\altaffilmark{2},
and Kazunari Shibata\altaffilmark{2}}

\altaffiltext{1}{Department of Astronomy, Kyoto University, 
Sakyo-ku, Kyoto 606-8502, Japan; kaorin@kwasan.kyoto-u.ac.jp}
\altaffiltext{2}{Kwasan and Hida Observatories, Kyoto University,
Yamashina-ku, Kyoto 607-8471, Japan}
\altaffiltext{3}{Department of Earth and Planetary Science, University of Tokyo, Hongo, Bunkyo-ku, Tokyo 113-0033, Japan}

\begin{abstract}
On 2005 September 13
a filament eruption accompanied by a halo CME
occurred in 
the most flare-productive active region NOAA 10808 
in Solar Cycle 23.
Using multi-wavelength observations
before the filament eruption on Sep.~13th, 
we investigate the processes leading to the catastrophic eruption.
We find that the
filament slowly ascended at a speed of $0.1 {\rm km \ s^{-1}}$
over two days before the eruption.
During slow ascending,
many small flares were observed 
close to the footpoints of the filament,
where new magnetic elements were emerging.
On the basis of the observational facts
we discuss the triggering mechanism leading to the 
filament eruption. 
%We suggest that
We suggest the process toward the eruption as follows: 
First, a series of small flares played a role in changing the 
topology of the loops overlying the filament.
Second, the small flares gradually changed the 
equilibrium state of the filament
and caused the filament to ascend slowly over two days.
Finally, a C2.9 flare that
occurred when the filament was close to 
the critical point for loss 
of equilibrium directly led to the catastrophic filament 
eruption right after itself.
\end{abstract}

\keywords{Sun: coronal mass ejections (CMEs) ---
Sun: filaments --- Sun: flares}

%%%%

\section{Introduction}
Investigating the triggering mechanism of filament eruption 
is one of the most important subjects in the field of space 
weather, because 
such ejections of plasma and magnetic fields from the sun 
lead to a significant disturbance
of the magnetosphere and affect human life. 
%%% Yurchyshyn et al. 2001 ApJ
It is widely accepted that 
flares, filament eruptions and coronal mass ejections (CMEs) are 
%attributed to the magnetic energy release 
different aspects of the same physical process 
including plasma ejection and magnetic energy release
(see \citealt{1995ApJ...451L..83S};
\citealt{2000JGR...10523153F}).
Several precursor signatures were reported 
\citep{1980SoPh...68..217M}.
However, what directly 
triggers such catastrophic eruptive events is not yet understood.
Many theoretical models were proposed so far 
(e.g., \citealt{1992LNP...399...69M},
\citealt{1999ApJ...510..485A}) 
%\citealt{2001ApJ...552..833M}).
As a catastrophe model for CMEs,
\citet{1990JGR....9511919F} and \citet{1991ApJ...373..294F}
proposed the loss of equilibrium.
Using a two-dimensional numerical simulation, 
\citet{1990JGR....9511919F} 
confirmed that when the filament current exceeds a 
critical value, the stable configuration containing the filament
loses equilibrium.
When equilibrium is lost, the filament is magnetically driven 
upward and may evolve into a CME.
\citet{2000ApJ...545..524C} and \citet{2001JGR...10625053L}
proposed the emerging flux triggering mechanism for filament 
eruptions and CMEs, motivated by the observed correlation
between reconnection-favored emerging flux and CMEs
 (e.g. \citealt{1995JGR...100.3355F}, 
\citealt{1999ApJ...510L.157W}).

From the observational point of view, however,
conclusions reported in literatures have not converged yet
in that which model is the best as the triggering mechanism of
eruptive events.
Therefore, accumulation of detailed analyses of
individual events is still necessary to understand 
the physical mechanisms of solar 
eruptions as well as to construct empirical models toward
space weather forecast.
In order to extract the essential physics and to compare with the 
existing theories, those events whose magnetic configuration
is as simple as possible may be suitable.
On the other hand, flare-productive active regions usually have
very complex magnetic structures.
From the viewpoint of space weather, we wish to understand
and predict the eruptions from such complex active regions.

Therefore, in this paper we focus on the eruptive events
in active region NOAA 10808 
which appeared in September 2005 and showed the
most furious flare activity during 
the current solar cycle (cycle 23). 
To have a clear view of the mechanism of the filament eruption 
in this region,
we examined extensively the observational signatures
that may lead to the filament eruption
using various available data.
We find several interesting features 
that might help to understand 
the physical mechanism of the eruption.
One of those is that the filament 
that erupted on Sep.~13th gradually deviated from the 
magnetic neutral line
over two days.
The speed of this ascending motion is approximately 
$0.1 {\rm km \  s^{-1}}$.
Another feature is the sites where C- and M-class flares occurred
during the period of ascending motion.
In these sites magnetic elements emerged and moved markedly.
%In this paper,
We discuss a plausible triggering mechanism of the eruption,
considering these features as the clues.
Independent analyses of this filament eruption
were carried out by 
\citet{2007Chifor} and \citet{2007Wang}.

In \S \ref{sec:NOAA10808}, we give 
an overview of active region NOAA 10808. 
We describe the observations of
a filament eruption that occurred on 2005 September 13
in \S \ref{sec:20050913X15}
and the long-time evolution of the filament in 
\S \ref{sec:filament_evolution}.
Discussions and conclusions are given in 
\S \S \ref{sec:10808Discussion} and 
\ref{sec:10808conclusion}, respectively.

\section{Overview of the active region}
\label{sec:NOAA10808}
\begin{figure}[htpb]
%\plotone{./GOES_lc_0907_17_1213waku.eps}
\plotone{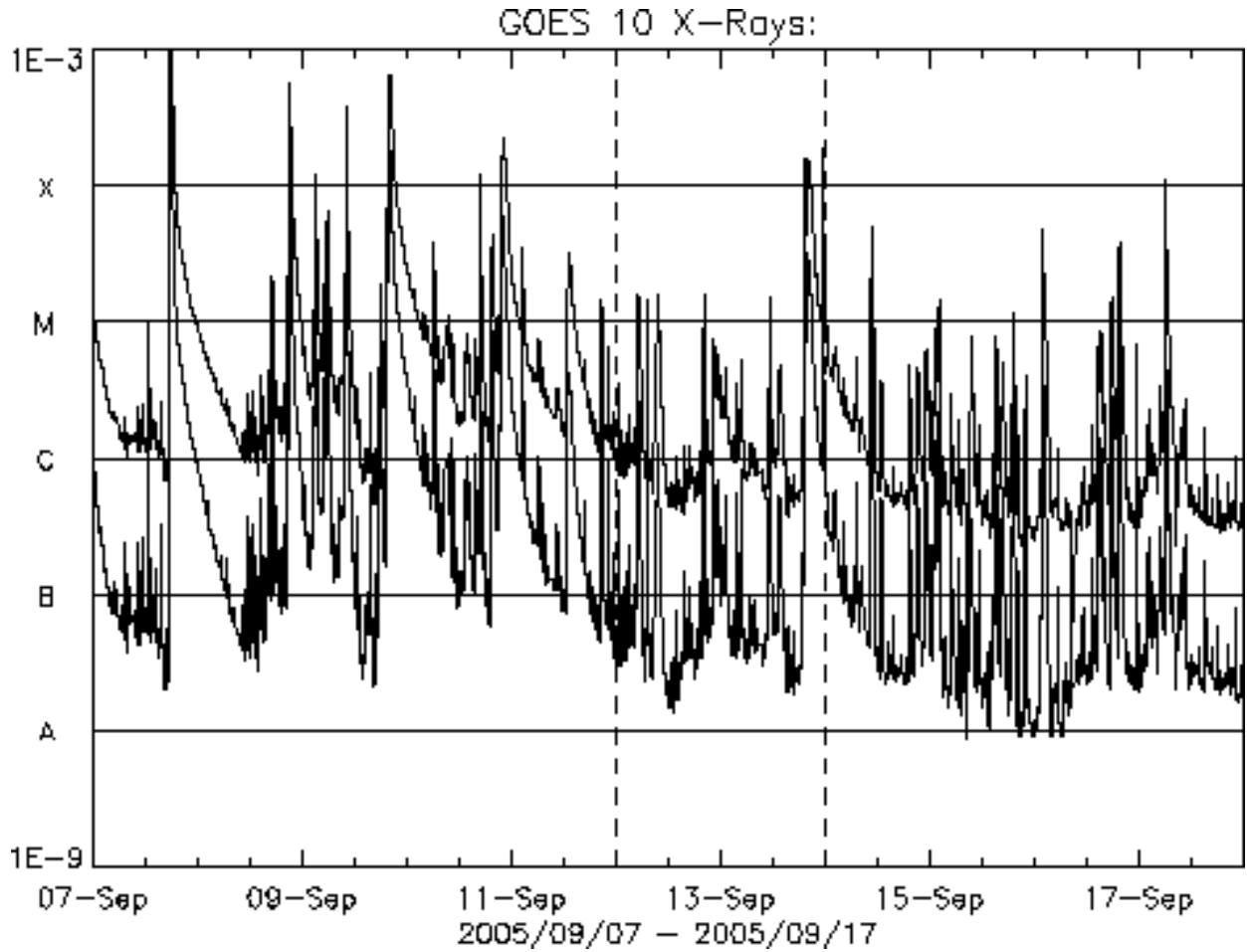}
\caption[{\it GOES} lightcurve from 7th to 17th in September 2005.]
{Soft X-Ray fluxes in the {\it GOES} 0.5-4 \AA \ (lower) 
and 1-8 \AA \ (upper) channels from 7th to 17th in September 2005.
The data was obtained by {\it GOES} 10.
In \S \ref{sec:preflare}, we focus on the M- and C-class flares 
that occurred  during the period between the dashed lines (12th and 13th).}
\label{fig:GOES_lc}
\end{figure}

In September 2005, a flare-productive 
active region appeared on the solar disk.
Active Region NOAA 10808 appeared over 
the east limb on 
2005 September 7 and produced at least 10 X-class 
and 25 M-class flares until it disappeared behind the disk.
Figure \ref{fig:GOES_lc} shows the {\it GOES} soft X-ray 
lightcurve of the period  when this active region was on the disk. 
Since no other strong active region was on the disk
during this period,
almost all the flares shown in this figure were attributed 
to this active region. This is the most intense flare 
activity during Solar Cycle 23 
despite in the declining phase of the cycle.
After crossing the meridian the active region seemed to 
decay remarkably and most of the X-class flares occurred 
in the eastern hemisphere.
Some flares were accompanied by CMEs; 
however, they were not so geoeffective.
(See \citealt{2006ApJ...646..625W}.)

It should be noted that
this active region was the return of active region NOAA 10798
of the previous month.
The active region NOAA 10798 emerged into a coronal hole
near the disk center
and formed a sea-anemone-type active region 
\citep{1994ApJ...431L..51S}.
By the time it rotated over the limb,
two M-class flares accompanied with halo CMEs occurred and 
caused a large geomagnetic storm with the minimum Dst 
index of $-216 {\rm nT}$ \citep{2006IAUJD...3E..72A}.
%(Asai et al. private communication).

\begin{figure}[htpb]
\begin{tabular}{ccc}
\includegraphics[scale=0.75]{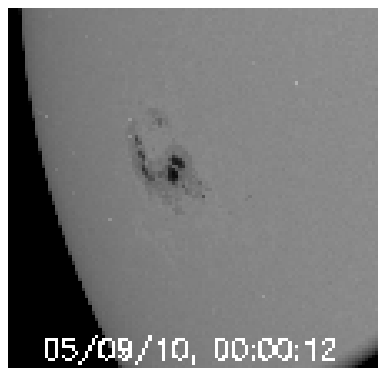} &
\includegraphics[scale=0.75]{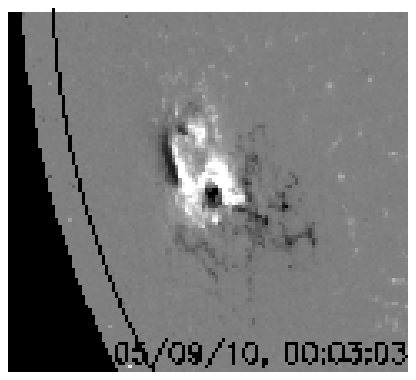}&
\includegraphics[scale=0.75]{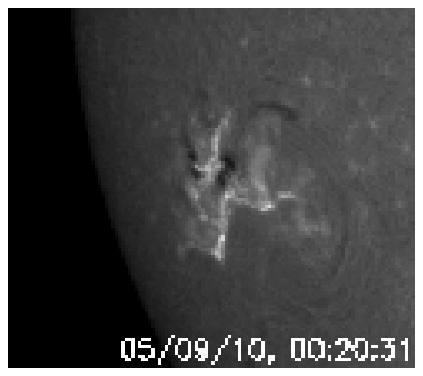}\\
\includegraphics[scale=0.75]{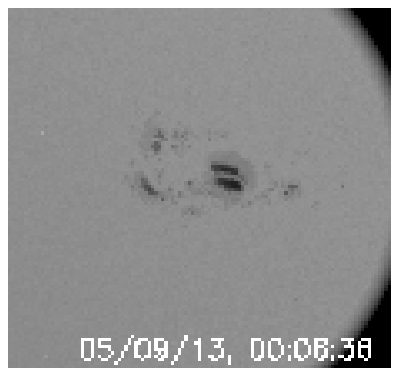} &
\includegraphics[scale=0.75]{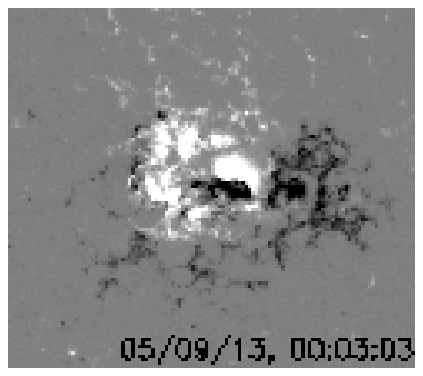}&
\includegraphics[scale=0.75]{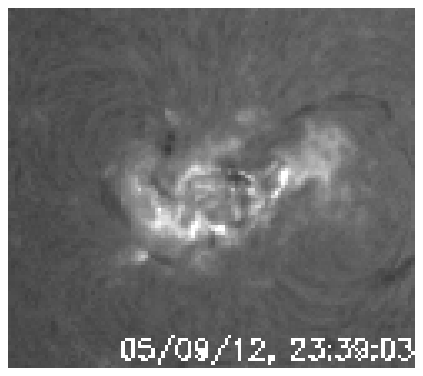}\\
\includegraphics[scale=0.75]{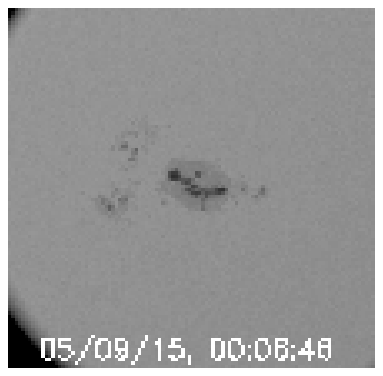} &
\includegraphics[scale=0.75]{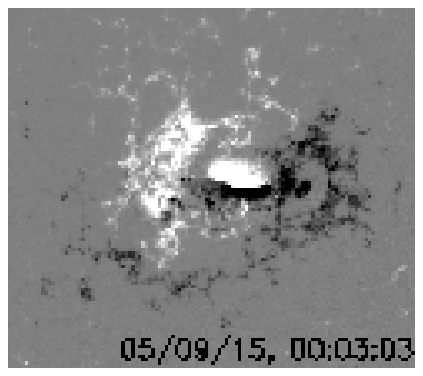}&
\includegraphics[scale=0.75]{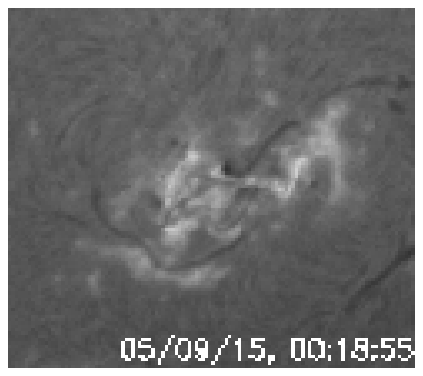}\\
\includegraphics[scale=0.75]{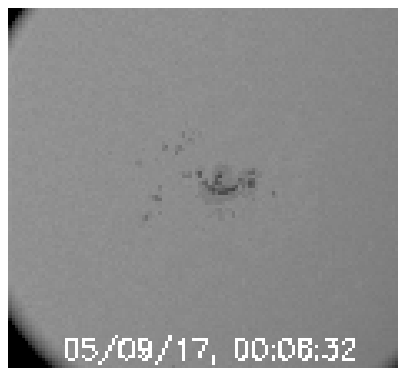} &
\includegraphics[scale=0.75]{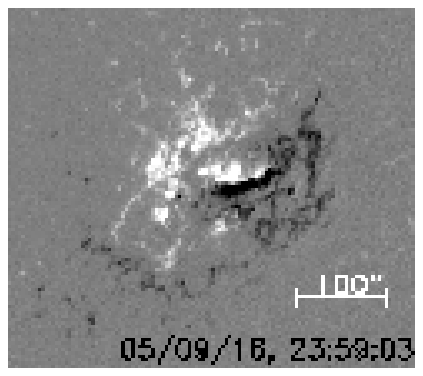}&
\includegraphics[scale=0.75]{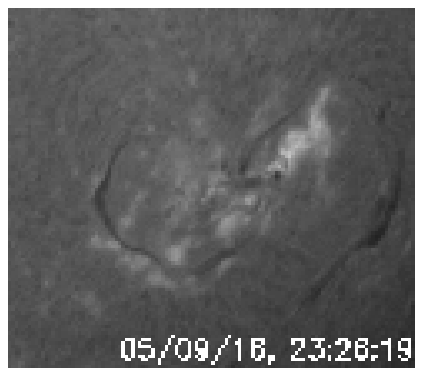}\\
\end{tabular}
\caption[Evolution of the active region NOAA 10808.]
{Evolution of the active region NOAA 10808.
{\it TRACE} white-light images, {\it SOHO} MDI magnetograms, and
SMART H$\alpha$ images are shown in the first, second,
and third columns, respectively.
In magnetograms, white and black indicate positive and
negative polarities, respectively, and 
solid lines indicate the solar limb.
The field of view of these images is $450'' \times 400''$.
North is to the top, and west is to the right in these and
other solar images in this paper.% Sterling & Moore 2005 ApJ Fig.1
}
\label{fig:AR_evolution}
\end{figure}

In Figure \ref{fig:AR_evolution},
four sets of images show
the evolution of the active region.
Each set has white-light image,  photospheric magnetogram,
and H$\alpha$ image co-aligned with each other.
The white-light images were taken by  
{\it Transition Region And Coronal Explorer}
({\it TRACE}; \citealt{1999SoPh..187..229H}).
The magnetograms are those from 
Michelson Doppler Imager 
(MDI; \citealt{1995SoPh..162..129S}) onboard 
the {\it Solar and Heliospheric Observatory} ({\it SOHO}).
The H$\alpha$ images were clipped
from the full-disk images obtained by 
Solar Magnetic Activity Research Telescope (SMART;
%\citealt{2004SPIE.5492..958U})
\citealt{2004ASPC..325..319U}) 
at Hida Observatory, Kyoto University.

At the center of the active region there was 
a delta-type sunspot which has two umbrae with 
different magnetic polarities packed tightly 
within a single penumbra.
It showed a 
counter-clockwise rotational motion.
The direction of the motion was such that
the orientation of the two polarities of the delta-type spot
approaches the so-called
%toward that fits to so-called 
``Hale's polarity law'' \citep{1919ApJ....49..153H}.
According to Hale's law, in southern hemisphere
the leading and following spots have
negative and positive polarities, respectively, 
during Solar Cycle 23.
The delta-type spot is surrounded by more diffuse magnetic field
that satisfies Hale's law.
Running through the delta-type sunspot,
there was an S-shaped neutral line in this active region
(see Figure \ref{fig:AR_evolution}; 
details are given in \S \ref{sec:20050913X15}).
Along the neutral line dark filaments were seen in H$\alpha$ images
in Figure \ref{fig:AR_evolution}.
We find that when halo CMEs occurred on Sep.~9th, 11th, and 13th,
the east part of the filaments erupted.
Although the other halo CME occurred on Sep.~10th,
we cannot identify the source region of the CME due to lack of 
EUV and H$\alpha$ data.
Among these events, we focus on the filament eruption 
that occurred on Sep.~13th to understand the 
mechanism of flares and
their relationship to the processes of eruptive events
in this active region.

\section{Filament eruption on 2005 September 13}
\label{sec:20050913X15}

On 2005 September 13, an X1.5 class flare occurred 
near the disk center in the super active region NOAA 10808.
This flare was accompanied by filament eruptions and 
a halo CME.
% ref. Zhang et al. 2001 APJL
A chronological description of the event is 
summarized in Table \ref{tab:050913X15}.
% ref. Chifor et al. 2006
Hereafter, this event is referred as ``the X1.5 event''.

\begin{figure}[htpb]
%\plotone{./ForRevise_images/GOES_lc_0913.eps}
\plotone{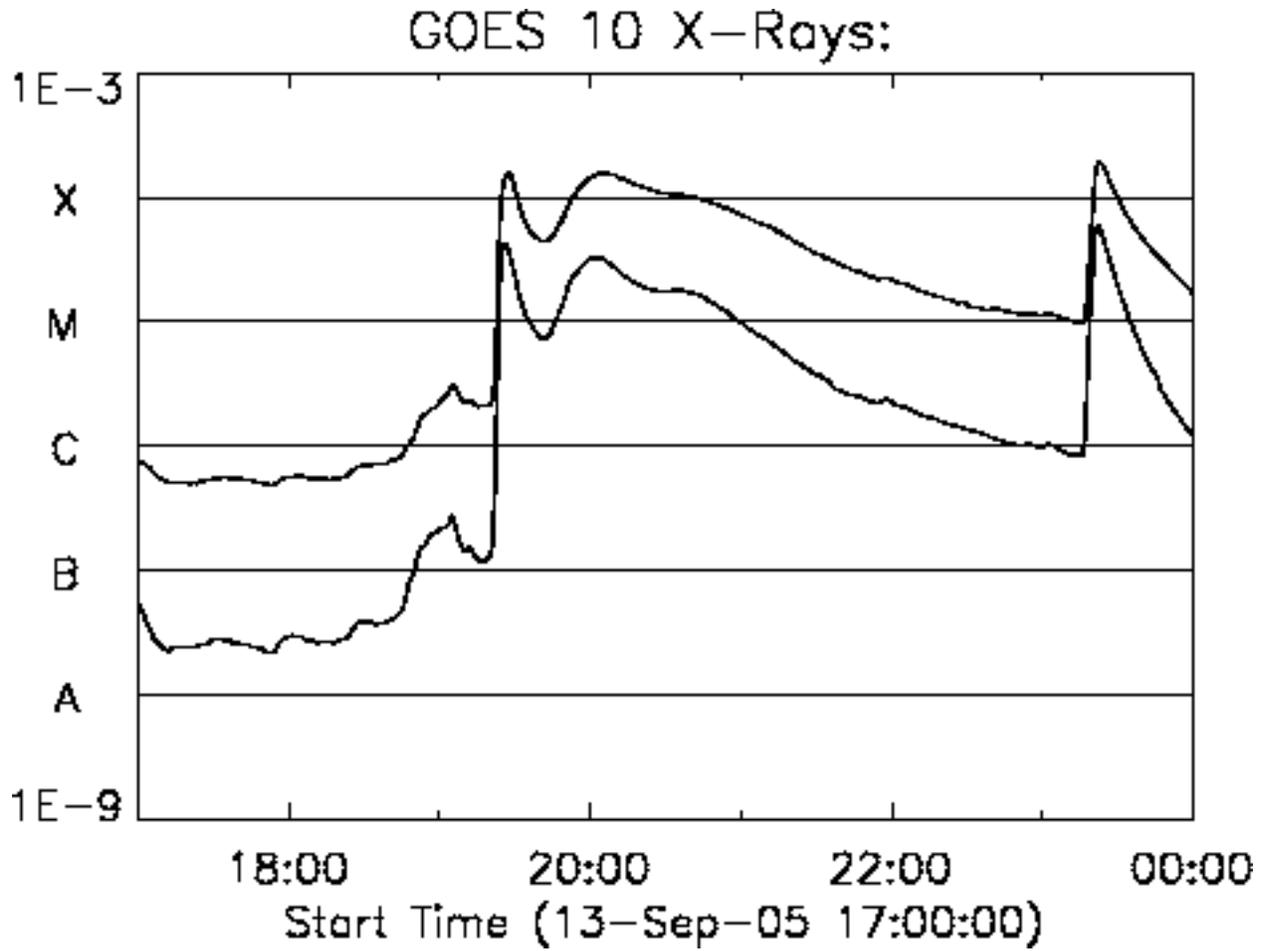}
\caption[{\it GOES} lightcurve on 2005 September 13.]
{Soft X-Ray fluxes in the {\it GOES} 
0.5-4 \AA \  (lower) and 1-8 \AA \ 
(upper) channels on 2005 September 13.}
\label{fig:GOES_lc_0913}
\end{figure}

\begin{figure}[htpb]
\begin{tabular}{ccc}
\includegraphics[width=0.3\textwidth]{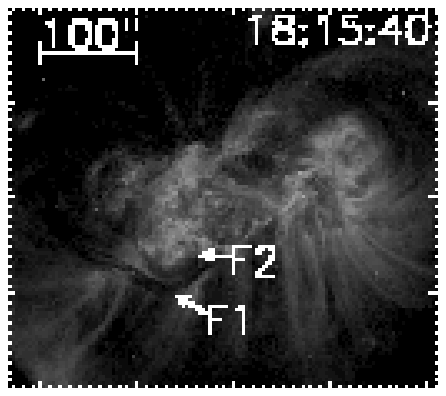}&
 \includegraphics[width=0.3\textwidth]{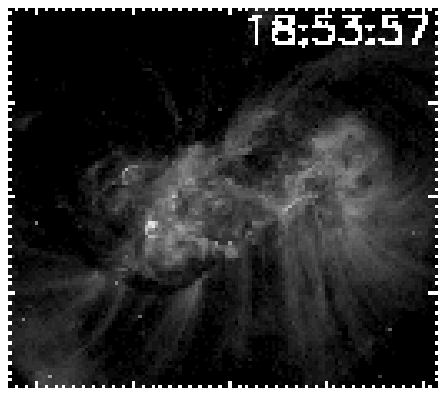}&
\includegraphics[width=0.3\textwidth]{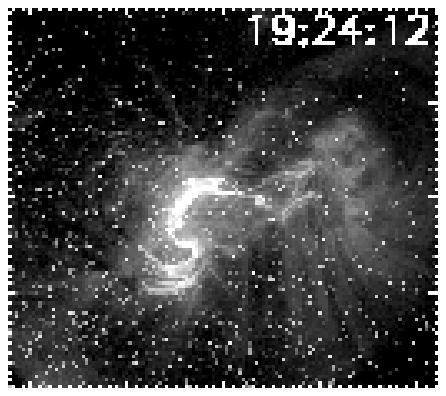} \\
 \includegraphics[width=0.3\textwidth]{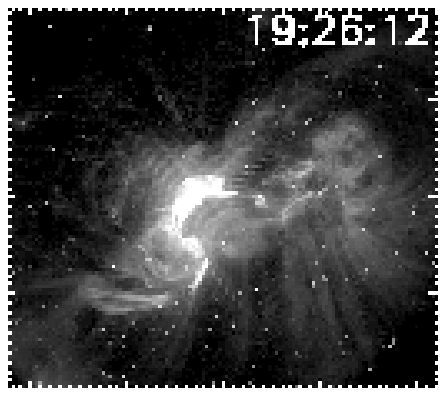}&
\includegraphics[width=0.3\textwidth]{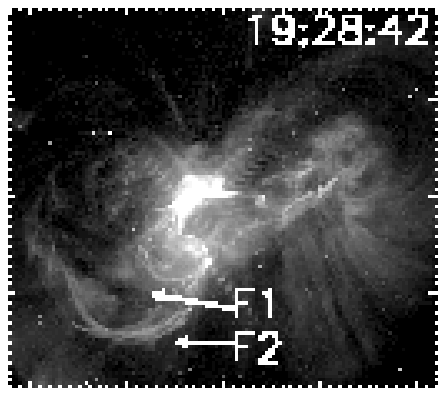} &
 \includegraphics[width=0.3\textwidth]{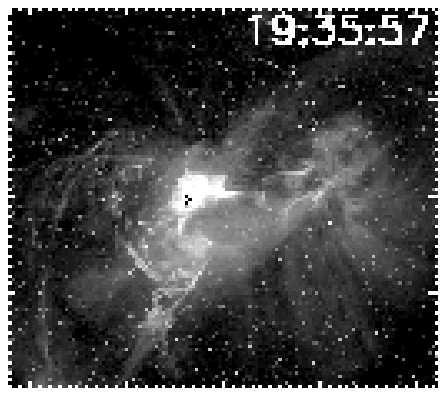} \\
\includegraphics[width=0.3\textwidth]{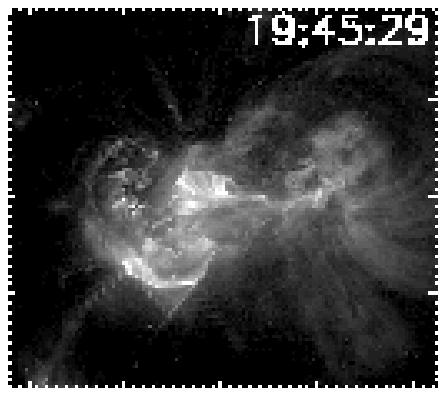} &
 \includegraphics[width=0.3\textwidth]{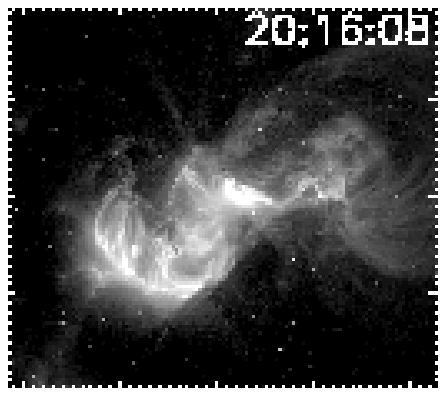} &
 \includegraphics[width=0.3\textwidth]{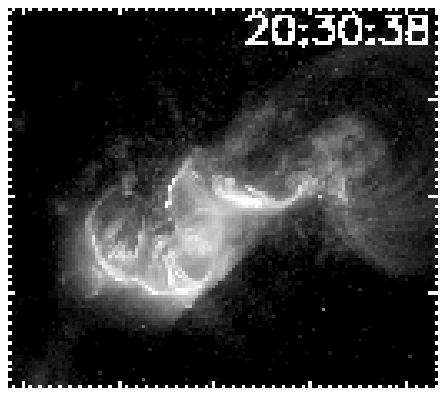}\\
\end{tabular}
\caption[EUV images of active region NOAA 10808 on Sep.~13th.]
{{\it TRACE} 195 \AA \  images of active region NOAA 10808 
before and during the X1.5 event on 2005 September 13.
F1 and F2 indicate filament 1 and 2, 
respectively (see text).
The field of view is $450'' \times 400''$,
the same as that of Fig.~\ref{fig:AR_evolution}
and the ticks correspond to $10''$.
This figure is also available as an 
mpeg animation in the electronic edition
of the {\it Astrophysical Journal}.}
% Mandrini et al. 2006 SoPh Fig.1}
\label{fig:TRACE195_0913X15}
\end{figure}

Figure \ref{fig:GOES_lc_0913} shows 
{\it GOES} soft X-ray light curves during this event.
The {\it GOES} 1-8\AA \ flux began to increase 
at 19:19 and attained its peak at 19:27.
% ref. Yurchysyn et al. 2006 SoPh
At the time of {\it GOES} flux peak, 
two filaments erupted: 
the inner one (hereafter referred as filament 2) 
brightened and erupted, 
the outer one (referred as filament 1)
remained dark and seemed to ascend following filament 2.
Figure \ref{fig:TRACE195_0913X15} describes the 
motion of the filaments as they appeared
in 195 \AA \ images 
% ref. Chifor et al. 2006 A&A
taken by {\it TRACE}.

\begin{figure}[htpb]
%\plotone{./ForRevise_images/eruption_ht_line_2.eps}
\plotone{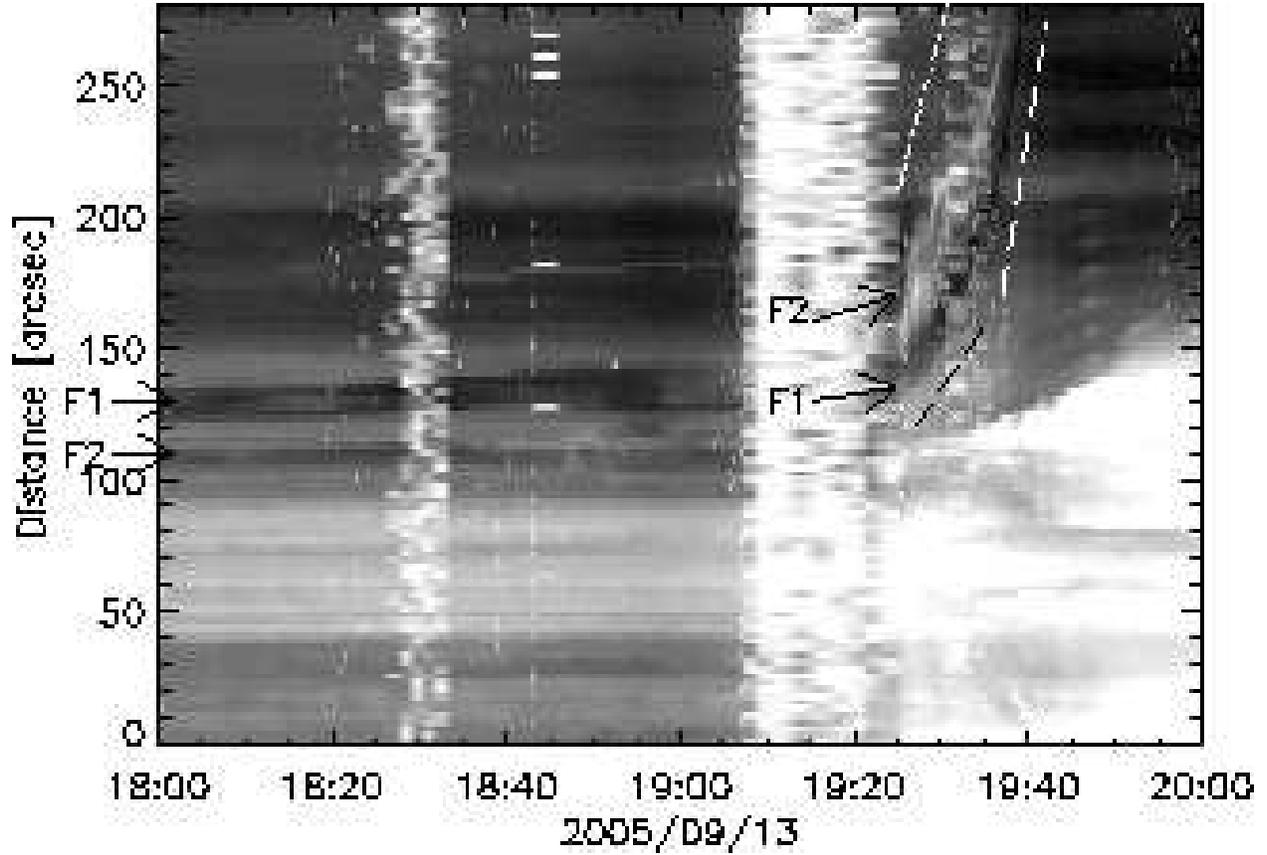}
\caption[Time slice showing the filament eruption.]{Time slice for slit shown in Figure 
\ref{fig:0913X15event}b. 
The spatial slices are 
laid adjacent from left to right in the diagram.
The distance is measured from 
the northwestern end of the slit.
%Isobe & Tripathi 2006 A&A Fig.2}
Arrows labeled as F1 and F2 point to filaments 1 and 2.
White dot-dashed line, white dashed line, and 
black dashed line indicate
$1.5 \times 10^2 \ {\rm km \ s^{-1}}$,
$2.5 \times 10^2 \ {\rm km \ s^{-1}}$,
and $5.8 \times 10 \ {\rm km \ s^{-1}}$, respectively.}
\label{fig:timeslice_filerup}
\end{figure}

\begin{figure}[htpb]
\begin{center}
\begin{tabular}{ccc}
\includegraphics[width=5.0cm]{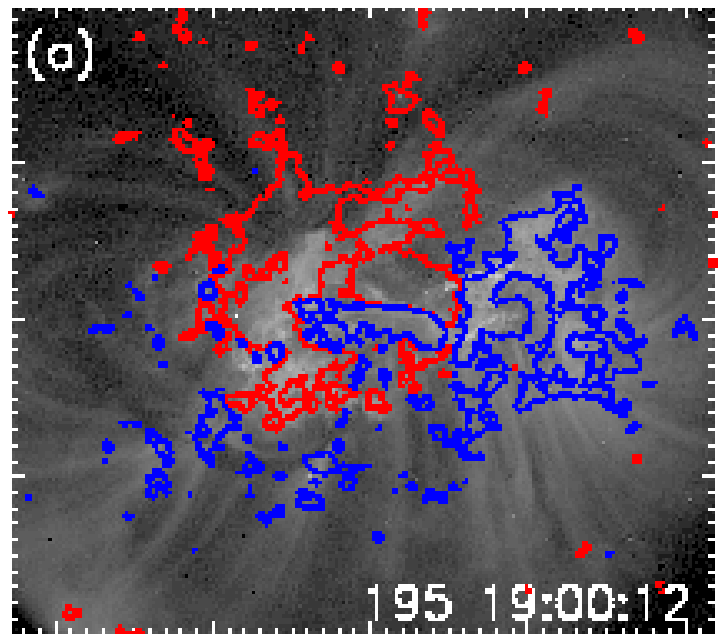}  &
\includegraphics[width=5.0cm]{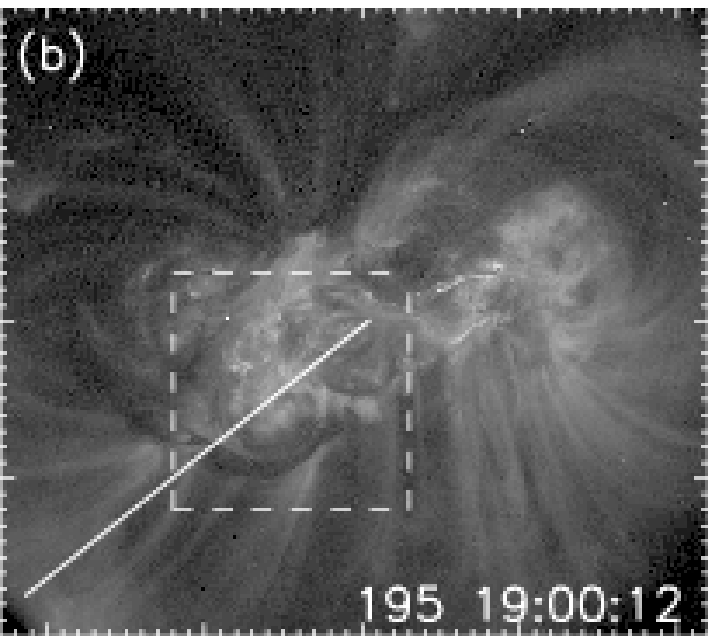} &
\includegraphics[width=5.0cm]{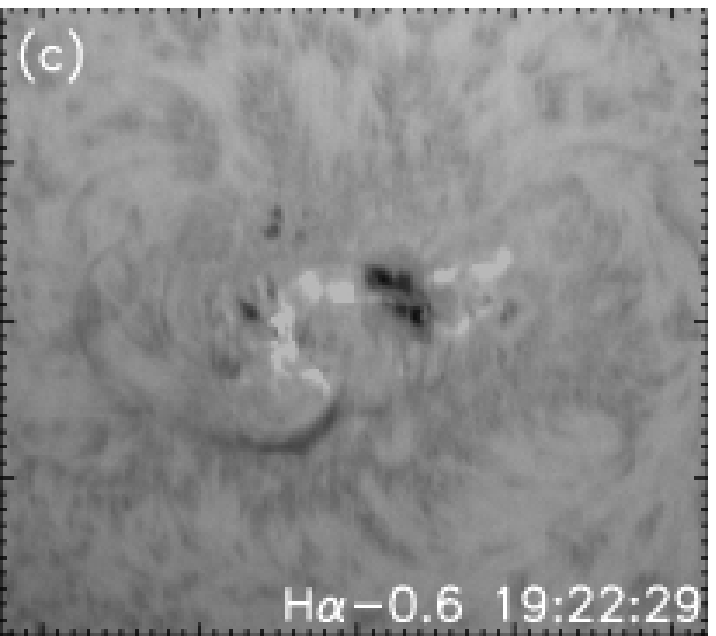} \\
\includegraphics[width=5.0cm]{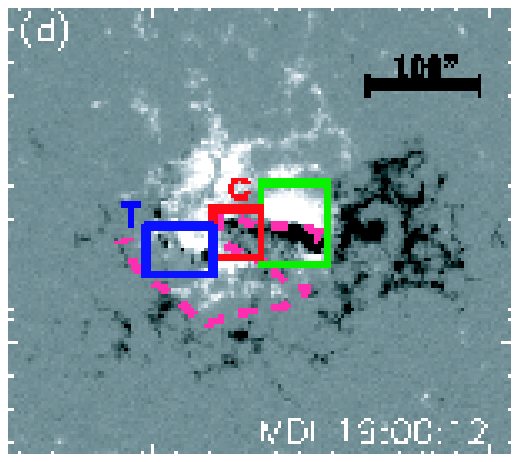} &
\includegraphics[width=5.0cm]{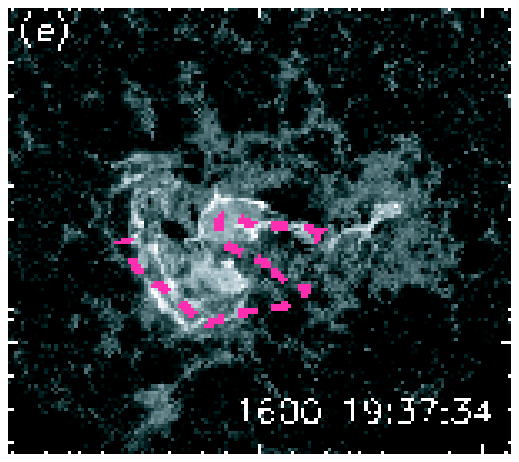} &
\includegraphics[width=5.0cm]{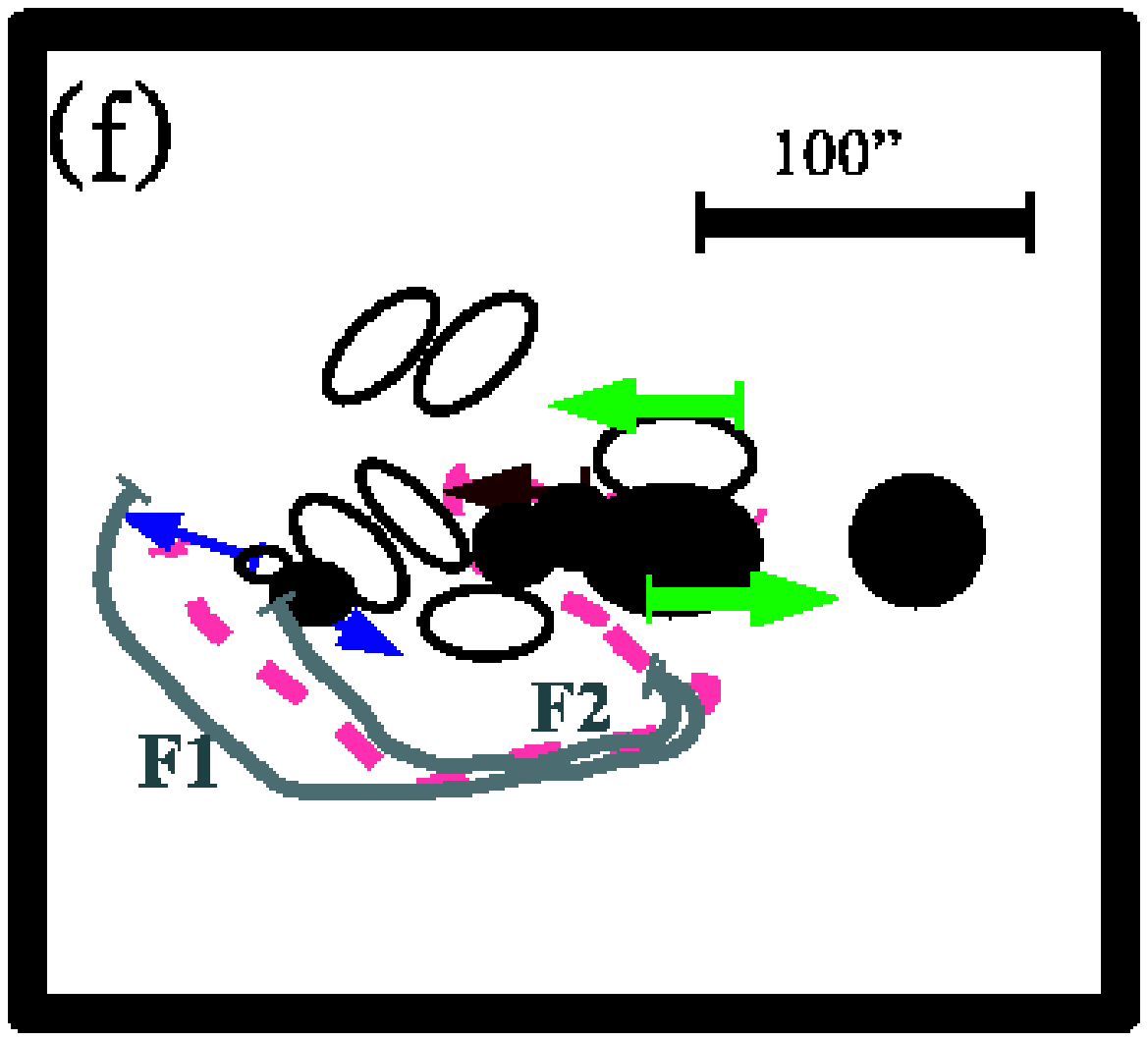} \\
\end{tabular}
\end{center}
\caption[Figures showing 
magnetic structure of the active region 
NOAA 10808.]
{Magnetic structure of active region 
NOAA 10808. The field of view of these images except 
Fig.~\ref{fig:0913X15event}f is $450'' \times 400''$,
the same as that of Fig.~\ref{fig:AR_evolution}.
A $100''$-long-scale that corresponds to about
$7\times10^{4} {\rm km}$ on the solar surface 
is shown in Fig.~\ref{fig:0913X15event}d.
(a){\it SOHO} MDI magnetogram contours superimposed on 
{\it TRACE} 195 \AA \  image.
{\it TRACE} 195 \AA \  image was taken at 19:00:12 UT 
and the magnetogram was taken at 19:11:03 UT. 
Displacement between these images due to difference in time and 
pointing is corrected. Red and blue contours represent 
$\pm 100 {\rm G}$.
(b){\it TRACE} 195 \AA \ image. The solid line indicates the slit 
position of time slice shown in 
Figs.~\ref{fig:timeslice_filerup} and \ref{fig:long_timeslice}.
%%Isobe & Tripathi 2006 A&A}
%The box with solid lines indicates the field of view 
%of a series of images in Fig.~\ref{fig:miniflare_mag195}.
The box written by dashed lines indicates the field of view 
%the area covered by 
of a series of {\it TRACE} 195 \AA \ images 
in Fig.~\ref{fig:miniflare_mag195} and of
a series of magnetic maps shown in Fig.~\ref{fig:MDI_preflare}.
(c)BBSO H$\alpha$ wing (H$\alpha -0.6$ \AA) image of the active 
region before the X1.5 event. 
%Brightening in EUV also began in region C at this time.
%%Yurchyshyn et al. 2006SoPh Fig.1
(d){\it SOHO} MDI magnetogram. A green box indicates 
the delta-type sunspot and a pink dashed line 
indicates the S-shaped neutral line. Region C in the red box and
region T in the blue box are the sites where small flares 
occurred frequently.
(e){\it TRACE} 1600 \AA \  image taken at 19:37:34 UT.
A pink dashed line indicates the S-shaped neutral line.
Flare ribbons appeared along the S-shaped neutral line.
(f)Schematic illustration of this active region.
In this panel,
the central part of the active region is enlarged.
White and black circles indicate positive and 
negative polarity regions, respectively. The meaning of 
pink dashed line is the same as in Fig.~\ref{fig:0913X15event}d.
Gray lines indicate filaments.}
\label{fig:0913X15event}
\end{figure}

Figure \ref{fig:timeslice_filerup} shows the height-time
profile for the erupting filaments along the slit shown in 
Figure \ref{fig:0913X15event}b %\ref{fig:timeslice_slit}
as observed from {\it TRACE} 195 \AA \ images.
This height-time profile shows 
filament 2 brightened and rapidly erupted 
at a velocity of $1.5 \times 10^2 \  \rm{km \  s^{-1}}$
(white dot-dashed line).
Filament 1 remained dark and seemed 
to follow bright filament 2 at a speed of
$5.8 \times 10 \ {\rm km \ s^{-1}}$ (black dashed line); 
the upward velocity increased up to 
$2.5 \times 10^2 \  \rm{km \ s^{-1}}$ (white dashed line).
Such motions of the filaments are also recognizable in 
Figure \ref{fig:TRACE195_0913X15}.

Figure \ref{fig:0913X15event} explains the magnetic structure of 
the active region and the locations of the filaments and the 
flares.
Figure \ref{fig:0913X15event}a %\ref{fig:TRACE195_mag_0913X15} 
shows the {\it SOHO} MDI magnetogram overlaid on the 
{\it TRACE} 195 \AA \  image before the eruption.
In this region,
there was an S-shaped neutral line %in this active region
running through the delta-type sunspot as shown in 
Figure \ref{fig:0913X15event}d.
The filament eruption 
occurred along all the southeast part of this neutral line.
At the same time of the eruption the flare core brightened in EUV 
on the east side of the delta-type sunspot
along the neutral line; this region is labeled as region C
(see Figure \ref{fig:0913X15event}d).
Region C is characterized by
several small negative-polarity magnetic elements moving 
one after another
from the negative-polarity umbra in the delta-type spot 
(see Figure \ref{fig:MDI_preflare} 
and \S \ref{section:regionC} for detail).
Another characteristic site in the magnetograms 
is on the opposite side of the neutral line from region C.
In this region, hereafter referred as region T, many
magnetic elements flowed out near the neutral line
(See Figure \ref{fig:MDI_preflare} and \S \ref{section:regionT} 
for detail).
As discussed in \S \ref{sec:preflare}, 
many small (C- and M-class) flares
occurred in regions C 
and T before the X1.5 event.
These characteristic magnetic regions 
are summarized in Figure \ref{fig:0913X15event}f
with their characteristic motions
indicated by arrows.
In this panel, filaments are indicated by gray lines:
Filament 1 (F1) originally appeared above the S-shaped neutral line
and gradually deviated southeastward, while filament 2 (F2)
appeared above the neutral line by the time the X1.~5 
event occurred.

Although the initiation of the filament eruption was unclear
in EUV because {\it TRACE} satellite 
was in the South Atlantic Anomaly (SAA) during the period, 
the first flare brightening in H$\alpha$ blue wing 
(H$\alpha -0.6$ \AA) 
was observed right before the eruption of filaments 1 and 2.
Figure \ref{fig:0913X15event}c
shows an H$\alpha$ wing (H$\alpha -0.6$ \AA) image
taken at the Big Bear Solar Observatory (BBSO).
The brightening began from region C as shown in Figure 
\ref{fig:0913X15event}c,
and expanded southeastward along the S-shaped neutral line.
After the filament eruption, 
flare ribbons formed along the 
S-shaped neutral line which are shown in {\it TRACE} 1600 
\AA \  images 
in Figure \ref{fig:0913X15event}e. %\ref{fig:TR1600_ribbon}.

After the filament eruption mentioned above,
a more faint eruptive event occurred in the south part 
of the active region at around 20:00 UT, i.e., at the second peak 
of {\it GOES} flux.
Footpoints of this erupting filament were located 
on the west of the delta-type sunspot and
close to the post flare loops in the southeast of
the active region.

These erupting events 
seemed to evolve into a white-light halo CME.
According to CME 
catalogue\footnote{http://cdaw.gsfc.nsa.gov/CME\_list}
\citep{2004JGRA..10907105Y},
this event is recognized as a halo CME
with the projected speed of $1866 {\rm km\ s^{-1}}$.
In {\it SOHO} LASCO C2 images, two blobs were observed;
the second bright core is thought to be related to the 
second eruption at 20:00 UT,
because its launch time is estimated at around 
that time on the basis of the linear extrapolation of its track.

\section{Long-time evolution of filaments}
\label{sec:filament_evolution}

\begin{figure}
%\plotone{./slowrise_ht_slitNL2.eps}
%plotone{./ForRevise_images/slowrise_ht_slitNL_arrow.eps}
\plotone{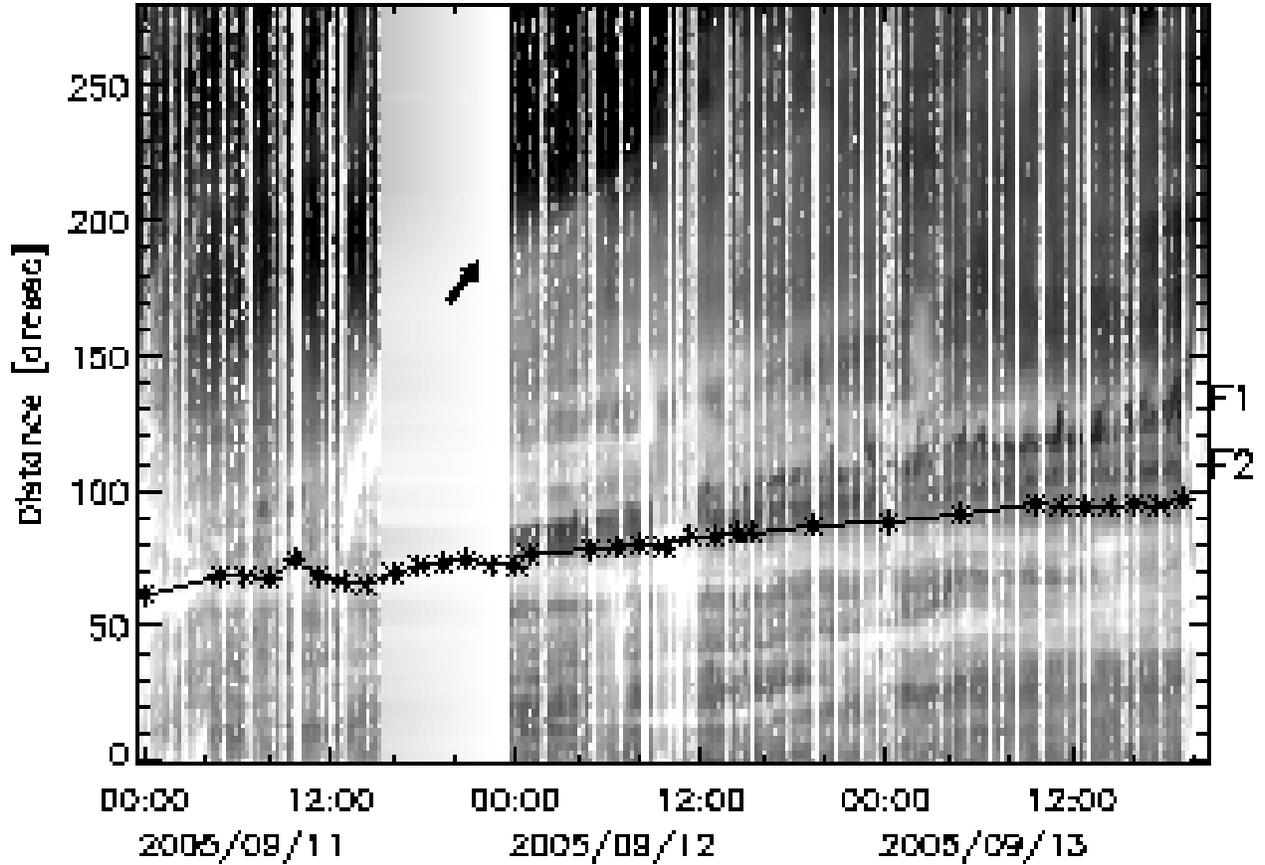}
\caption[Time slice showing slow ascending of the filament.]
{Time slice for slit shown 
in Fig.~\ref{fig:0913X15event}b.
The distance is measured from the northwestern end of the slit.
Asterisked line indicates the position of the 
magnetic neutral line.
Asterisks are the data points.
The error of the neutral line position is roughly 10 arcsecs. 
F1 and F2 indicate dark filaments 1 and 2, respectively.
The black arrow indicates the rising motion of a bright structure
(see \S \ref{sec:10808Discussion}).}
\label{fig:long_timeslice}
\end{figure}

Before Sep.~13th eruption (the X1.5 event), 
we can see two filaments 
in the southeast part of the active region:
The outer one (filament 1) 
and the inner one (filament 2).
They formed after another filament eruption on Sep.~11th.
We concentrate on the evolution of these filaments
after the eruption on Sep.~11th.
We find that filament 1 continuously deviated from the neutral
line from 11th to 13th.
Figure \ref{fig:long_timeslice} shows the time evolution of 
EUV intensity along the slit shown in Figure
\ref{fig:0913X15event}b.
Asterisked line indicates the position of the magnetic 
neutral line.
This figure shows that the filament seemed to 
separate from the neutral line at a 
nearly constant speed of $0.12 {\rm km \ s^{-1}}$ 
over more than
40 hours, while the filament accelerated to 
$200 - 300 {\rm km \  s^{-1}}$ when it erupted
as mentioned in \S \ref{sec:20050913X15}.
Although we see only the plane-of-sky motion of the
filament separating from the neutral line,
we can interpret it as an ascending motion
by taking into account the projection effect. 
Such a slow and long-lasting ascending motion is 
%not the motion during so-called slow-rise phase of
probably different from
so-called slow-rise phase of the erupting filaments,
such as those reported by \citet{1997PASJ...49..249O}
(see also 
\citealt{1988ApJ...328..824K},
\citealt{1990HvaOB..14...37R},
\citealt{2001ApJ...559..452Z},
\citealt{2004ApJ...613.1221S}).
\citet{2006A&A...458..965C} reported the early evolution of 
a prominence eruption observed on 2005 July 27.
Slow-rise at a speed of $4.8 {\rm km \  s^{-1}}$ was observed 
over 30 minutes prior to the fast-rise phase during which
the filament accelerated up to $300 {\rm km \  s^{-1}}$.
Since the slow ascending motion continues much longer than
those in the previously reported events,
we believe the physical mechanism is different.
\citet{2006A&A...449L..17I} and
\citet{2007preprint...Isobe} reported 
an observation of slow rising motion ($1 {\rm km s^{-1}}$) 
of a polar crown filament that last for more than 10 hours 
before it finally erupted. 
They found a large amplitude oscillation of the filament 
during the slow rising, and hence concluded that 
the filament retained a stable equilibrium and 
the rising motion is a quasi-static evolution. 
This may be the same phenomenon as our long-lasting 
slow rise.
Existence of such long-term slow rise has been
also pointed out by S. Martin (private communication).

It should be noted that
filament 1 showed active motion along its axis
in {\it TRACE} EUV images and SMART H$\alpha$ images
during its slowly ascending, i.e., over two days
(e.g., \citealt{1980SoPh...68..217M}).

\section{Preflare brightenings on Sep.~12th and 13th}
\label{sec:preflare}
\sectionmark{Preflare brightenings}

\subsection{All M- and C-class flares}

\begin{figure}[htpb]
\plotone{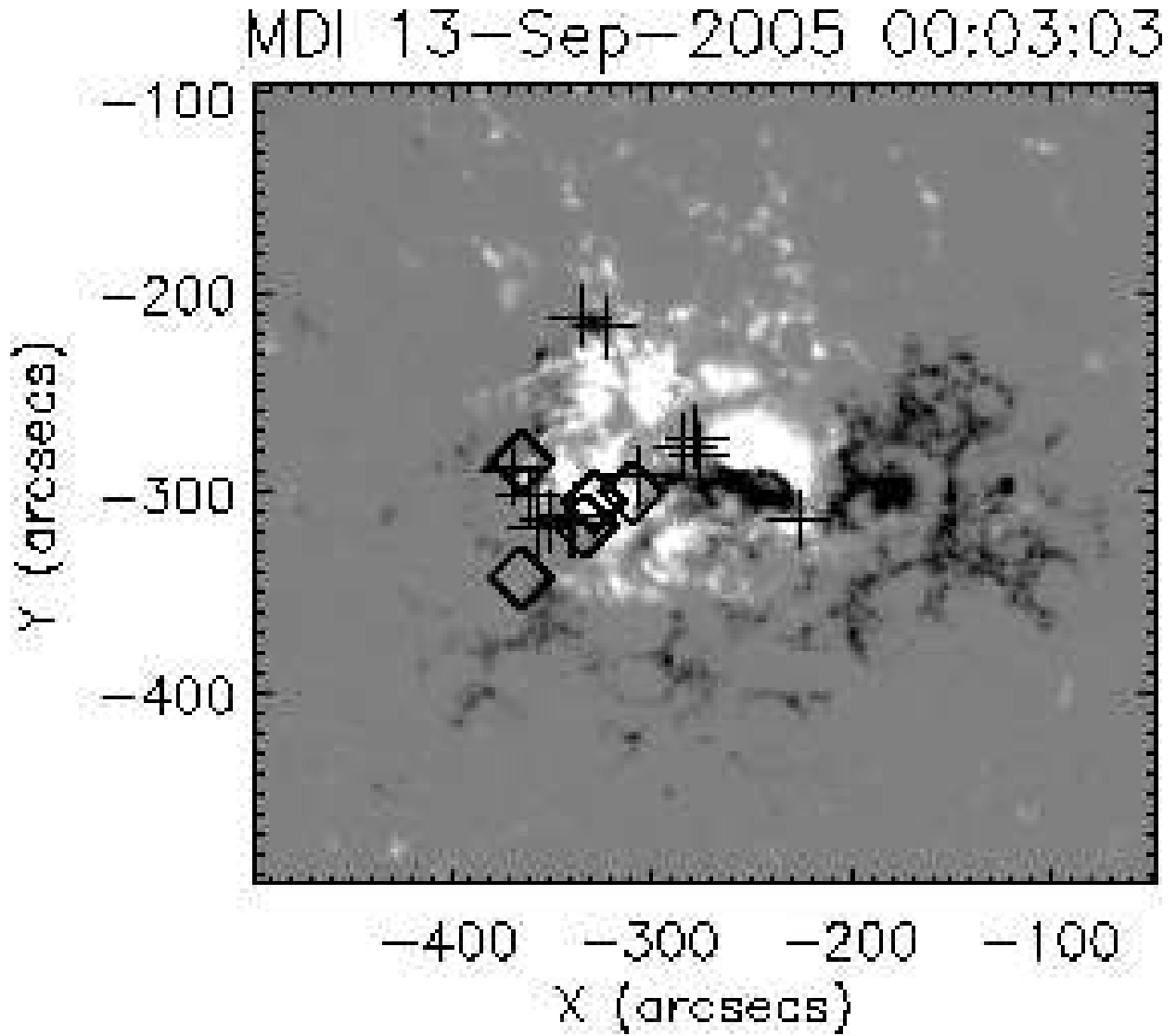}
%\plotone{./MCflares_ichi.eps}
\caption[Location of the M- and C-class flares on 2005 Sep.~12th and 13th.]
{Location of the M- and C-class flares that occurred on 
2005 September 12 and 13 showing on 
{\it SOHO} MDI magnetogram.
White and black indicate positive and negative polarities in the 
magnetogram, respectively.
Diamonds and crosses indicate M-class and C-class, respectively.
FOV is the same as that of Fig.~\ref{fig:AR_evolution}.}
\label{fig:MCflares_ichi}
\end{figure}

\begin{figure}[htpb]
\begin{tabular}{ccc}
%M1.5 (region C[group II]) & 
%M1.3 (region T[group III]) & 
%C1.5 (jet[group IV])\\
\includegraphics[width=0.3\textwidth]{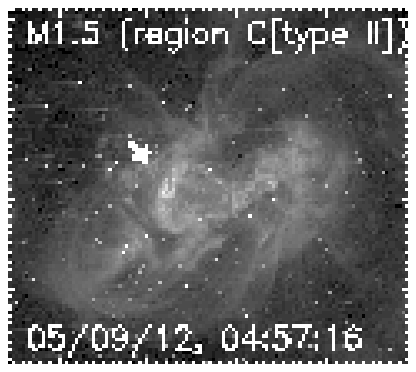} &
\includegraphics[width=0.3\textwidth]{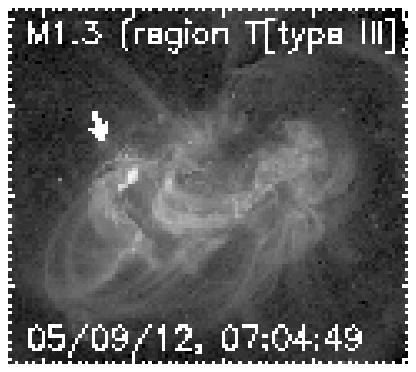} & 
\includegraphics[width=0.3\textwidth]{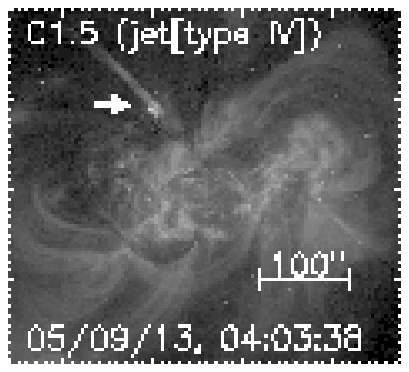} \\
\end{tabular}
\caption[EUV images of M- and C-class events.]
{{\it TRACE} 195 \AA \  
images showing representative events of group II, III, and IV
on 2005 September 12 and 13. 
Arrows indicate the sites of brightening.
FOV is the same as that of Fig.~\ref{fig:AR_evolution}.}
\label{fig:MCflares}
\end{figure}

During the slow rise of the filament,
several M- and C-class flares and small brightenings in EUV
occurred in this active region.
As shown between the dashed lines in Figure \ref{fig:GOES_lc},
5 M-class and 18 C-class flares occurred since 00:00 UT on 
Sep.~12th until the X1.5 flare on Sep.~13th.
All of these flares are listed in Table \ref{tab:MCclassflares}
and the locations on the magnetogram %of them are shown 
are shown in Figure \ref{fig:MCflares_ichi}.
These flares can be categorized into four groups:
(I) eruptive events along the S-shaped neutral line
(the X1.5 event),
(II) brightenings around filaments 
 on the east side of the delta-type spot (region C),
(III) local events in the emerging flux region  (region T),
(IV) small jets in the northeast 
part of the active region.
Since small jets (group IV) were localized events and
were not attributed to the S-shaped neutral line,
we consider that these events 
did not have direct influence on the X1.5 event.
{\it TRACE} 195 \AA  \  images of groups II, III, and IV
are shown in Figure \ref{fig:MCflares}.

Group I is the filament eruption 
during the X1.5 event on Sep.~13th itself.
We consider the other smaller events (groups II and III)
which occurred along the S-shaped neutral line
as a key to find out the triggering mechanism of the X1.5 event.
First, we focus on the events occurred in region C (group II)
to consider the eruptive process of the X1.5 event.
Next, we consider the last local event (group III) right before the
X1.5 event to consider the direct triggering mechanism of the 
X1.5 event.

\subsection{Filaments in region C and M1.5 and M6.1 flares on Sep.~12th}
\label{section:regionC}

One of the puzzling points in the X1.5 event is that
although the flare ribbon extended throughout the S-shaped 
neutral line, the filament was visible 
only in the southeast part of the neutral line.
We suggest that an eruptive process did occur in region C
as well, even though we did not observe an filament during 
the X1.5 event. The magnetic topology may be either 
a flux rope (e.g., \citealt{1989ApJ...344.1010P};
\citealt{1996SoPh..167..217L})
or a sheared arcade (e.g., \citealt{1994ApJ...420L..41A}).

\begin{figure}[htpb]
\begin{tabular}{ccc}
\includegraphics[scale=1.1]{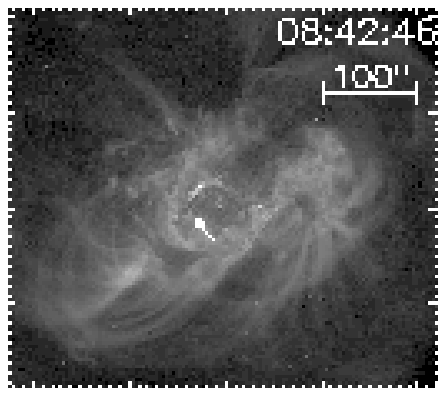} 
\includegraphics[scale=1.1]{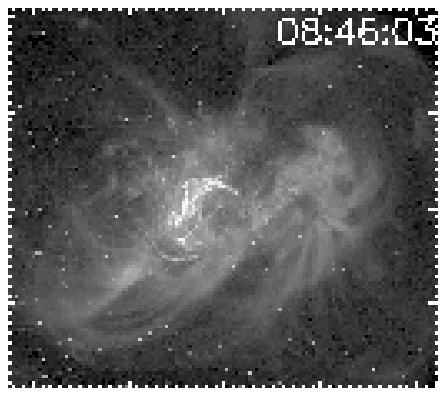} &
\includegraphics[scale=1.1]{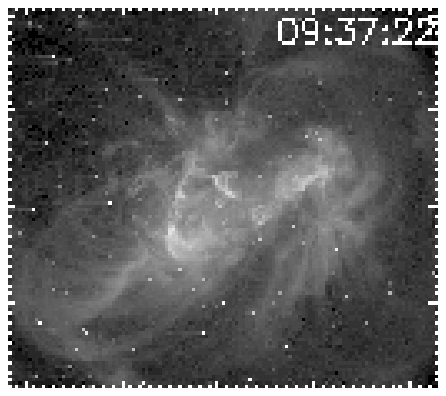} \\
\\
\\
\end{tabular}
\caption[EUV images showing the filament 
eruption in region C.]
{
{\it TRACE} 195 \AA \ images 
showing the filament eruption in region C.
During the M6.1 class flare at 9:03 UT on 12th,
the thin dark filament (arrow in the left panel) 
in region C appears to be lifted up by 
a brightening loop and erupted.
FOV of these images is the same 
as that of Fig.~\ref{fig:AR_evolution}.
}
\label{fig:regionCfil}
\end{figure}

A filament was actually visible in region C sometimes
during Sep.~12th and 13th.
An M1.5 flare occurred in region C at 5:05 UT on 12th.
An EUV brightening seemed to be
located under a dark filament in region C.
Neither eruption nor a CME was 
observed during this event.
Some loops overlying the dark filament 
brightened after the main phase.
Four hours after the M1.5 flare,
an M6.1 class flare occurred at 9:03 UT.
Figure \ref{fig:regionCfil} shows
{\it TRACE} 195 \AA \  images of this event.
In this event, brightening in EUV started from region C.
A thin dark filament in region C 
was lifted up %by the brightening loop 
and seemed to erupt at 8:46 UT, i.e., 
the first peak of the GOES soft X-ray flux.
The thin dark filament twined around the brightening loop.
When the brightening loop expanded the dark filament 
erupted. 
Although the dark filament could not be identified clearly after 
its eruption, according to CME catalog
\citep{2004JGRA..10907105Y},
a narrow CME was observed by {\it SOHO} LASCO 
right after this event.
A little thin lump of bright coronal emission 
first appeared in LASCO/C2 field of view at 9:12 UT 
% Zhou et al. 2006 p.1241
and was recognized as a CME with the projected speed 
of $511 \rm{km \  s^{-1}}$,
which is consistent with the assumption 
that the filament erupted at around 8:46 UT and evolved into a CME.

After these flares, a
dark filament existed in region C.
This filament could be seen in SMART H$\alpha$ wing 
(H$\alpha \pm  0.5 {\rm \AA}$)
images taken at around 3:00 UT on 13th. %by SMART T1. %;
We observed ceaseless mass flow along the filament 
in region C.
This is similar to the active motion of the filament 1 along 
its axis that is mentioned in \S \ref{sec:filament_evolution}.

On the basis of the observational features of 
the small dark filament in region C, 
we speculate that there was a flux rope 
or sheared field in region C.
Owing to the nonsteady mass flow along the filament
in region C, the filament itself was sometimes 
visible and, another time, was invisible.
However, the magnetic field that supported the filament
should have existed all the time.
Considering the fact that during the X1.5 event
the brightening in H$\alpha$ started 
from region C and extended to the southeast part, 
we suppose that 
this eruptive process in region C 
was the initiation of the X1.5 event
even though the filament in region C was invisible
at the time of the event.
Those observations described in this section 
support that a flux rope (or whatever plays the role of a filament)
did exist in region C even at the time of the X1.5 event on 13th. 

\subsection{Possible direct trigger for 
 the X1.5 event: C2.9 flare on Sep.~13th}
\label{section:regionT}
A small brightening in EUV close to the footpoint of 
filament 2 was observed 
40 minutes before the filament eruption.
It was recorded as a {\it GOES} C2.9 class flare at 19:05 UT
(see Figure \ref{fig:GOES_lc_0913}).
We consider this event as the final trigger of the X1.5 event.

\begin{figure}[htpb]
\begin{tabular}{ccc}
\includegraphics[scale=0.8]{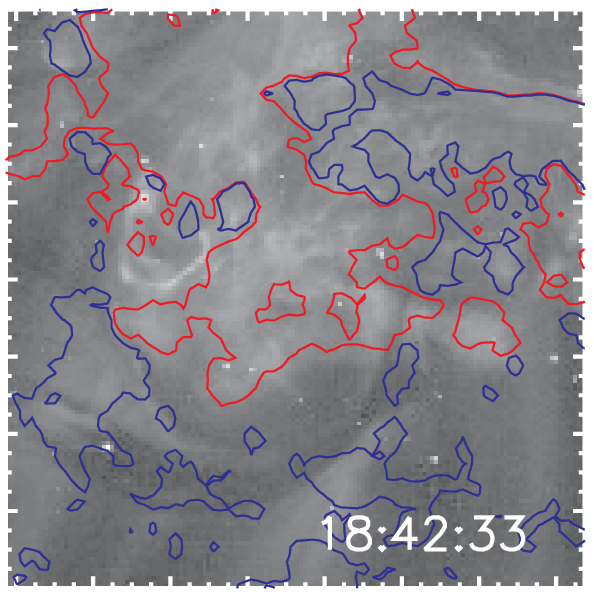} &
\includegraphics[scale=0.8]{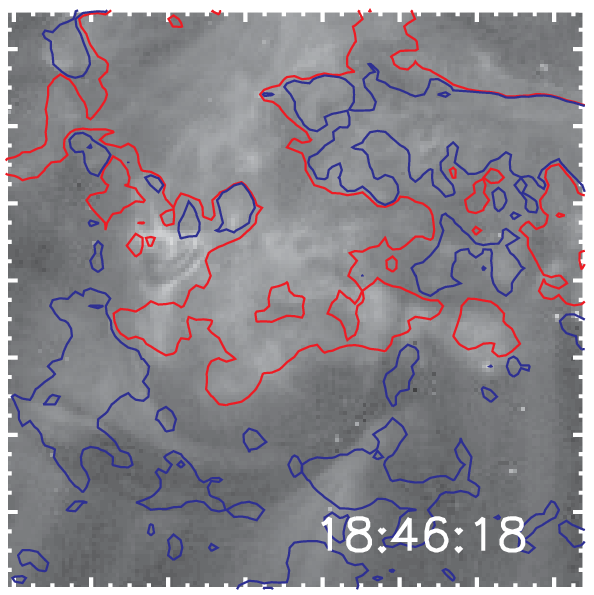}&
\includegraphics[scale=0.8]{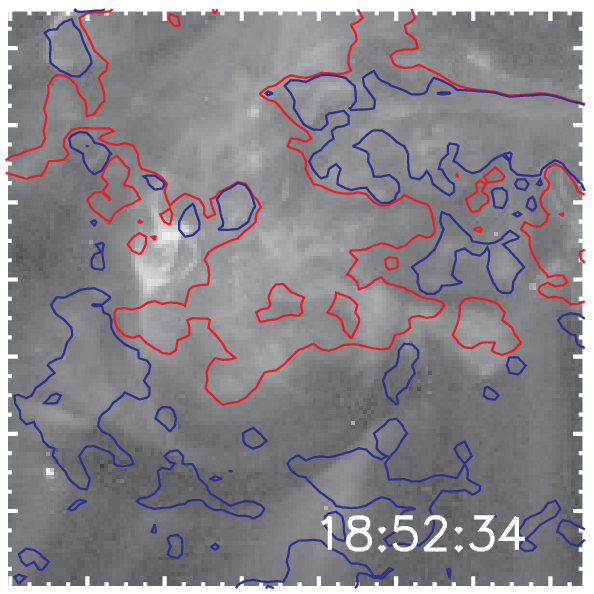} \\
\includegraphics[scale=0.8]{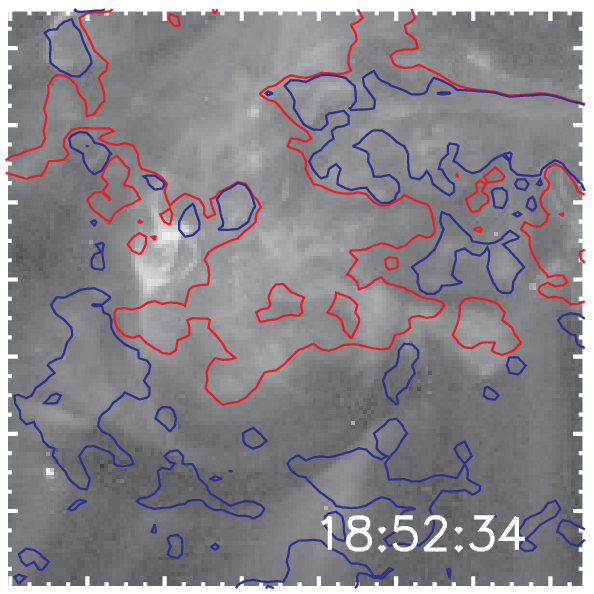} &
\includegraphics[scale=0.8]{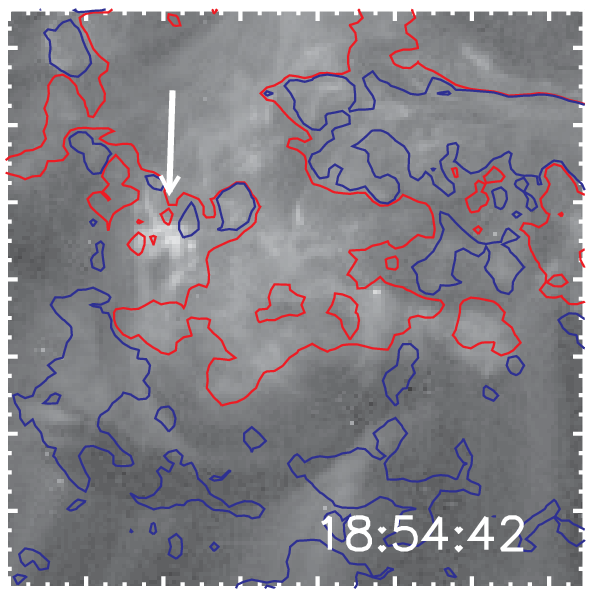}&
\includegraphics[scale=0.8]{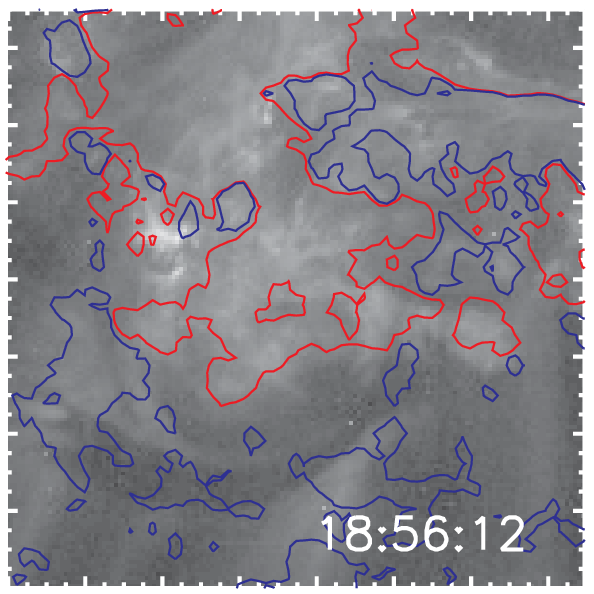} \\
\includegraphics[scale=0.8]{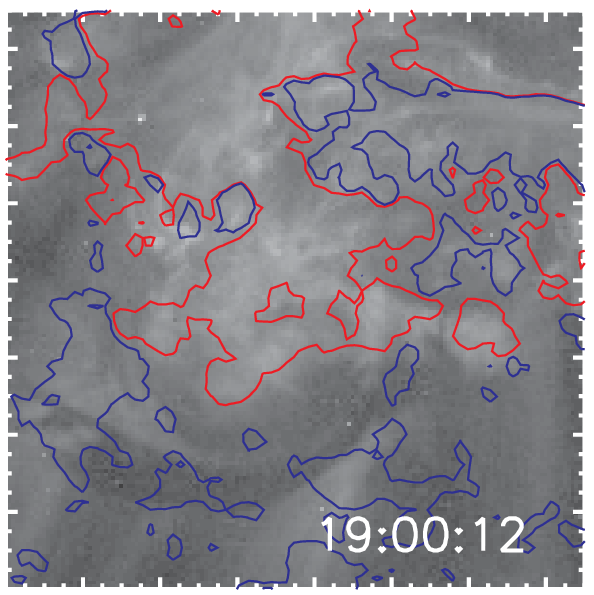} &
\includegraphics[scale=0.8]{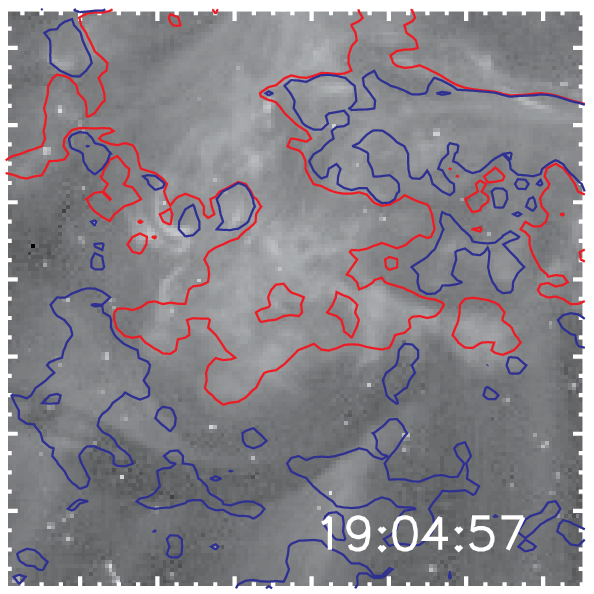}&
\includegraphics[scale=0.8]{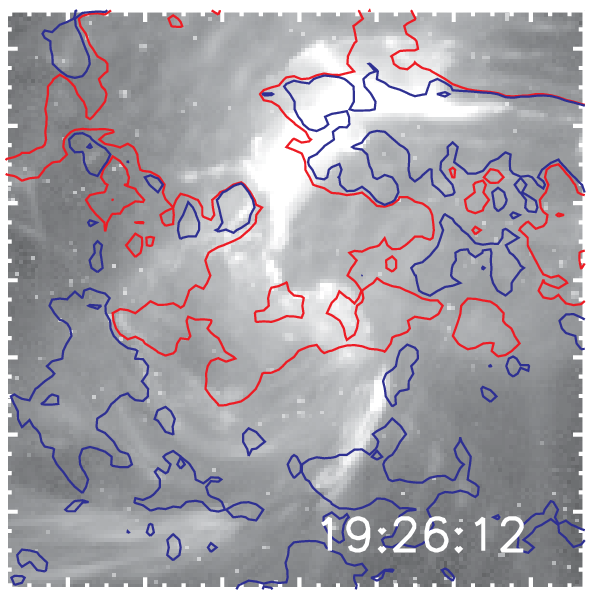} \\
\end{tabular}
\caption[MDI magnetogram and {\it TRACE} 195 \AA \  images of the C2.9 flare.]
{{\it SOHO} MDI magnetogram contours overlaid on 
{\it TRACE} 195 \AA \  images showing the C2.9 flare  
considered as a trigger of the X1.5 event.
The white arrow in the center panel indicates the jet-like feature.
The last panel shows the filament eruption 
during the X1.5 event (see Fig.~\ref{fig:TRACE195_0913X15})
to indicate the relative position against the C2.9 flare.
Red and blue contours indicate $+100 {\rm G}$ and $-100 {\rm G}$,
respectively.
The field of view of these images is shown in 
Fig.~\ref{fig:0913X15event}b
as a box with dashed lines and is $150''$ square.
}
\label{fig:miniflare_mag195}
\end{figure}

\begin{figure}[htpb]
\begin{tabular}{cccc}

\includegraphics[scale=1.0]{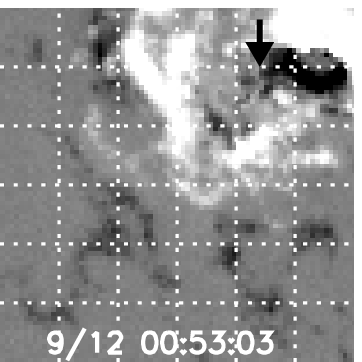}&
\includegraphics[scale=1.0]{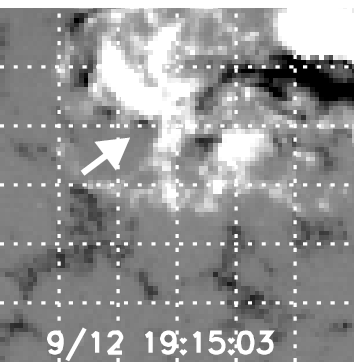}&
\includegraphics[scale=1.0]{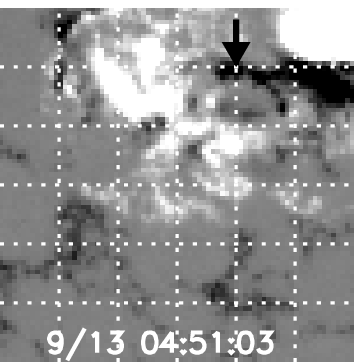} &
\includegraphics[scale=1.0]{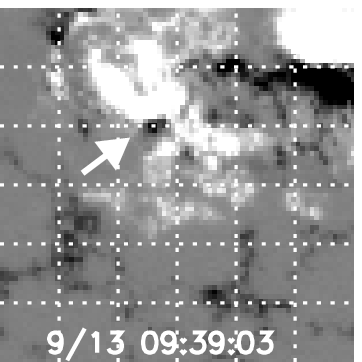} \\
\includegraphics[scale=1.0]{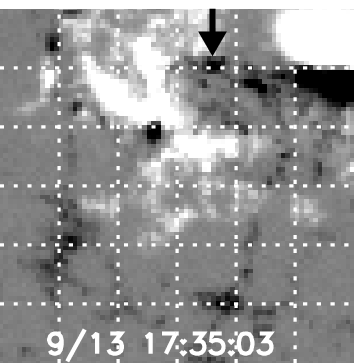}&
\includegraphics[scale=1.0]{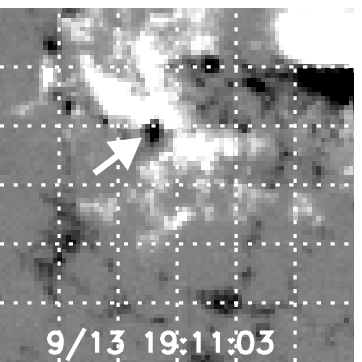} & 
\includegraphics[scale=1.0]{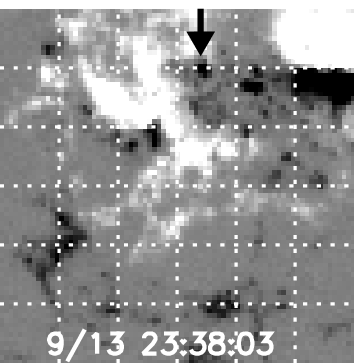} &
\includegraphics[scale=1.0]{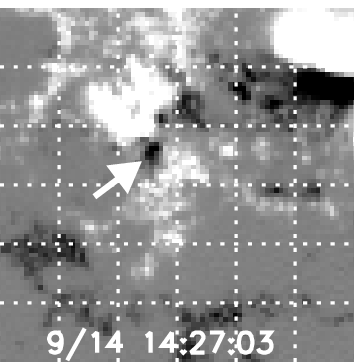}\\
\end{tabular}
\caption[Evolution of the 
magnetic field around the S-shaped neutral line.]
{{\it SOHO} MDI magnetic maps showing the evolution of the 
magnetic field around the S-shaped neutral line.
The area covered by each image is 150 arcsecs square 
which corresponds to the square drawn in 
Figs.~\ref{fig:0913X15event}b.
Dotted lines are drawn every 25 arcsecs 
($1.8\times 10^4 {\rm km}$).
Black arrows %in the first column
indicate the negative elements in region C 
flowing out from the negative umbra in the delta-type spot,
while white arrows % in the second column 
indicate the negative elements emerging into
region T and moving toward the footpoint of the loop
that brightened along with the C2.9 flare at around 19:00 UT 
on Sep.~13th. }
%Mandrini et al. 2006, Fig.1
\label{fig:MDI_preflare}
\end{figure}

At the site of the C2.9 flare,
successive appearances of small magnetic elements were seen 
in magnetograms.
Figure \ref{fig:miniflare_mag195} shows a series of 
{\it TRACE} 195 \AA \  images during this event.
{\it SOHO} MDI magnetogram contours are overlaid;
red and blue contours indicate $\pm 100 {\rm G}$.
At around the center of each panel, a small brightening in EUV
appeared and jet-like features pointed by the white arrow 
were observed to the east of the 
bright loops. 
These dark jet-like features are also 
seen in BBSO H$\alpha$ wing (H$\alpha - 0.6{\rm \AA}$) images.
These jets and loops moved eastward gradually and
the jets also showed a whip-like motion.
Moreover, on the left side of bright jet-like features, 
dark backward flows were observed when we check 
a {\it TRACE} 195 \AA \ movie of this event
(see mpeg animation of Figure \ref{fig:TRACE195_0913X15}).
Figure \ref{fig:MDI_preflare} shows a series of 
MDI magnetograms
mainly taken on September 12th and 13th.
From the tail part of the S-shaped neutral line,
lots of small magnetic elements were flowing out.
Although region T was dominated by the negative polarity,
patchy magnetic elements with both polarities were emerging.
Over the two days before the filament eruption,
similar small brightenings in EUV which were identified as 
C- or sub-C-class flares by {\it GOES} 
occurred in the similar locations (Figure \ref{fig:MCflares_ichi});
C-class events were listed in Table \ref{tab:MCclassflares}.
Taking a closer look at the sites of small brightenings,
they slightly moved southward with time, but
corresponded to the source region of 
the emerging mixed polarities.

It is noteworthy that an element of negative polarity 
which first appeared in the frame taken at 19:15 UT on Sep 12th
(see the second panel in the first row in 
Figure \ref{fig:MDI_preflare})
moved southwestward along the neutral line.
It weakened the positive region 
so that the positive region divided into parts
after the X1.5 event (see the magnetogram 
image taken at 14:27 on Sep.~14th in 
Figure \ref{fig:MDI_preflare}).
This negative element was located at the footpoint of the 
brightening loop during the C2.9 event at 19:05 UT on Sep.~13th.
We examine the motion of the negative element and
it yields valuable clues to understand the triggering mechanism 
of the X1.5 event
(see \S \ref{sec:10808Discussion}).

It should be noted that 
small brightenings in EUV 195 \AA \  were also found  
just before the X1.5 event on the west side of the delta-type spot.
In H$\alpha$ wing (H$\alpha-0.6 {\rm \AA}$)
images taken at BBSO, the only data taken
simultaneously during this period,
we cannot identify corresponding brightenings.
Although this event occurred in the 
same period as the C2.9 event 
we consider that this west-part event did not have 
a direct influence in triggering the X1.5 event.
The reasons are as follows:
(1)In this region similar brightenings in EUV were also observed 
one hour before this event.
(2)The dark filament in this region seemed not to be 
affected by the X1.5 event, and remained as it were
after the X1.5 event.
(3)Although it located close to the footpoint of the second 
eruption at around 20:00 UT, the magnetic field lines 
in this region did not seem to directly connect with 
those which were involved in the X1.5 event.
Another brightening was observed to the northeast 
of the tail of the S-shaped neutral line 
(See the center panel in the first row of 
Figure \ref{fig:TRACE195_0913X15}).
We consider that this event was not directly related to 
the X1.5 event because of similar reasons. %in a similar way.

\section{Discussion: Triggering Mechanism of the X1.5 Event}
\label{sec:10808Discussion}

On the basis of the multi-wavelength observations,
we discuss the triggering mechanism of the X1.5 event.
As mentioned in the previous section, 
we considered M- and C-class flares on Sep.~12th and 13th 
as a key to find out the triggering mechanism of the X1.5 event.

% filament formation (2)
A filament eruption occurred at 13:12 on Sep.~11th
accompanying an M3.0 flare.
At least ten hours after this filament eruption,
another filament was re-formed along the S-shaped neutral line.
After that several M- and C-class flares and 
small brightenings in EUV occurred in region C and region T.
We consider that these small flares played a role in 
lengthening the magnetic arcades overlying the filaments.

\begin{figure}[htpb]
\begin{tabular}{ccc}
\includegraphics[width=0.32\textwidth]{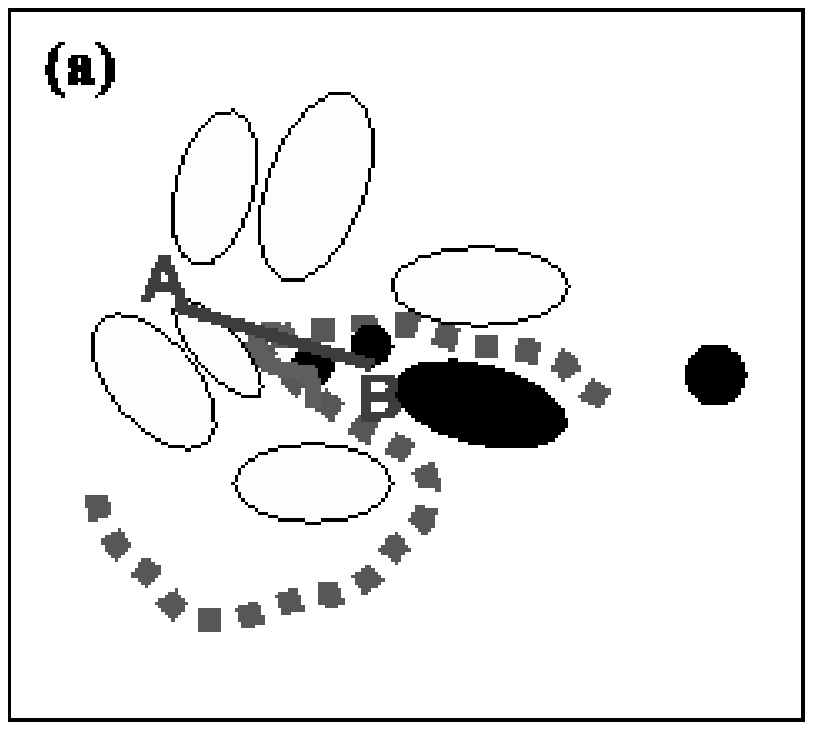} &&\\
\includegraphics[width=0.32\textwidth]{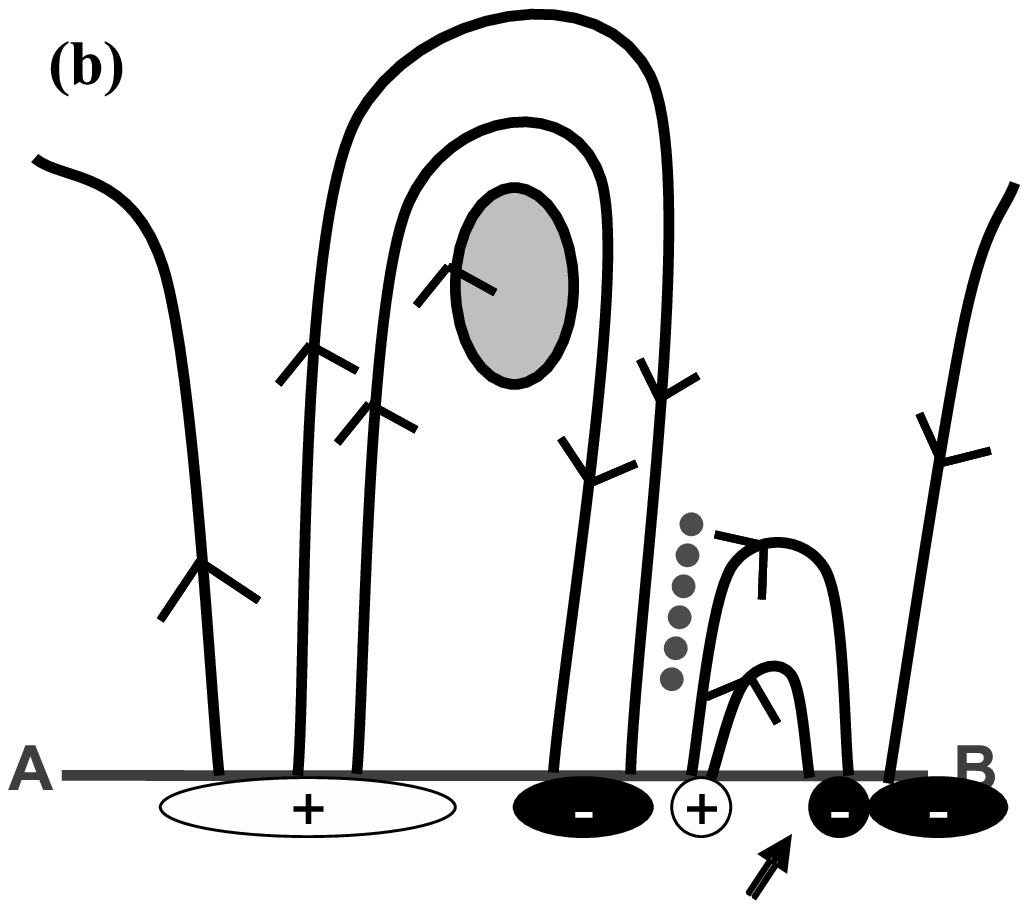} &
\includegraphics[width=0.32\textwidth]{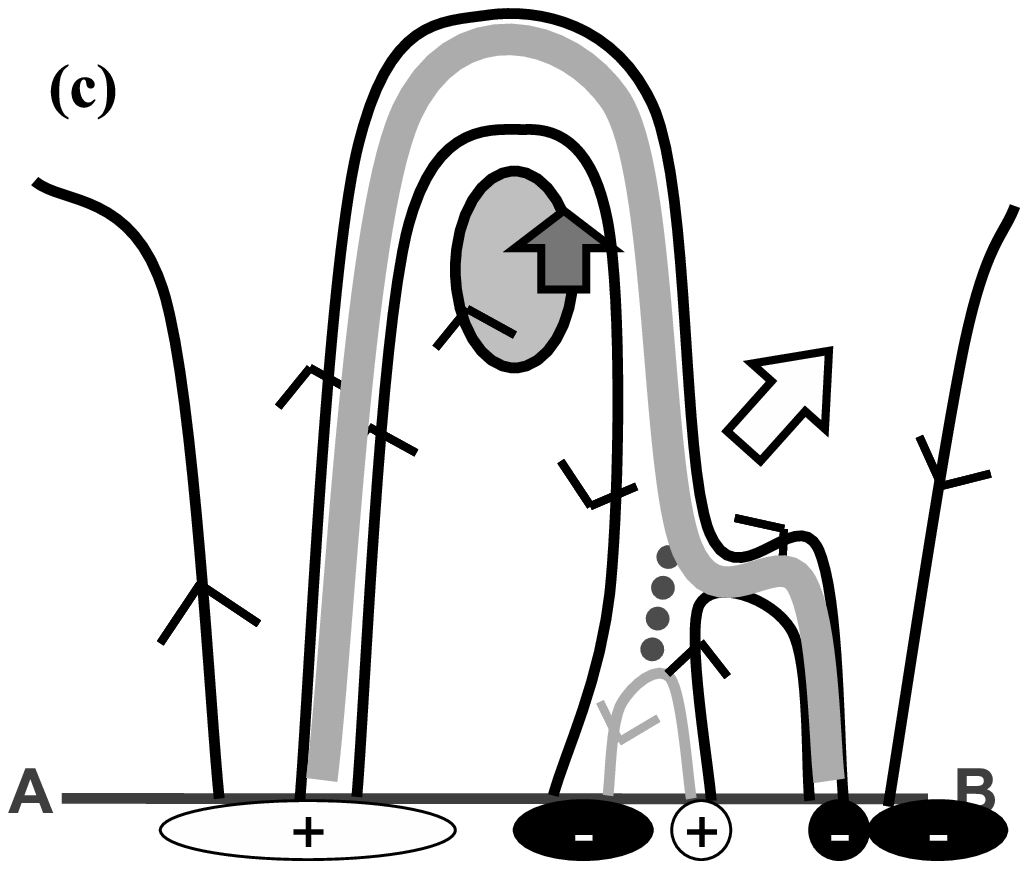} &
\includegraphics[width=0.32\textwidth]{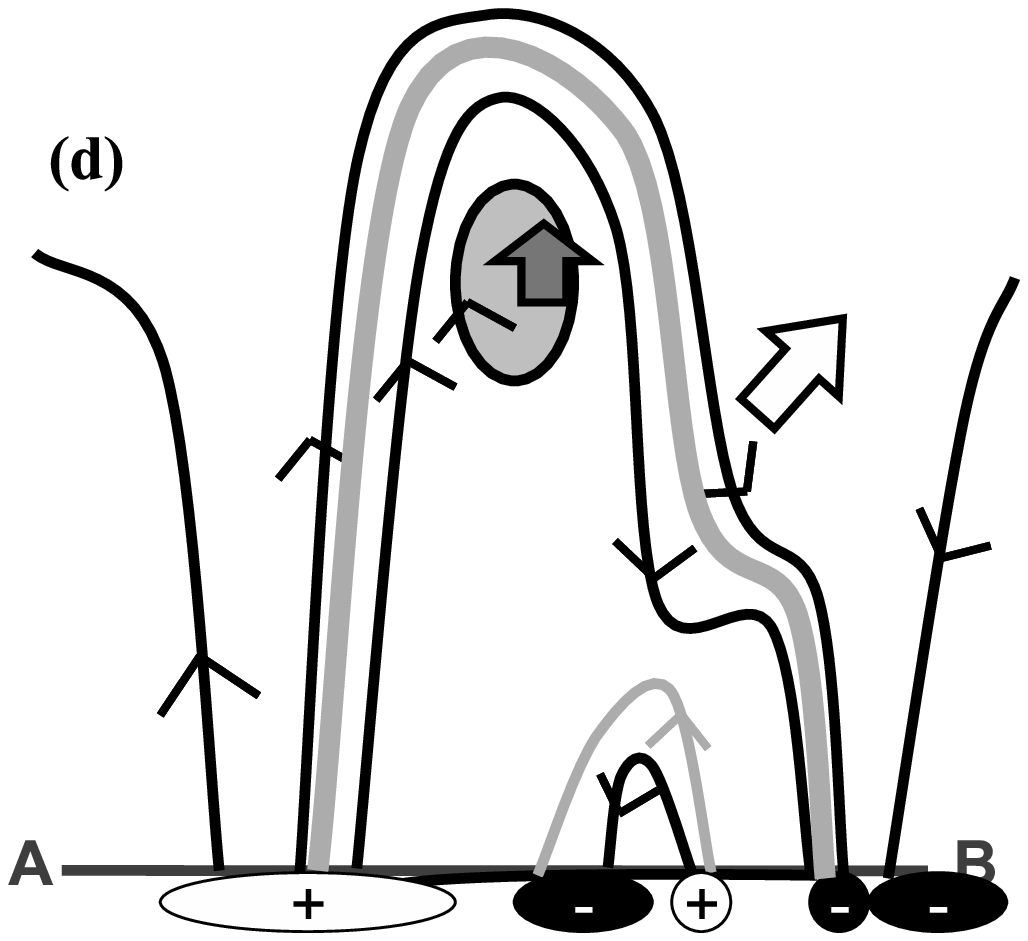} \\
\end{tabular}
\caption[Evolution of magnetic structure in region C. (the M6.1 flare)]
{Schematic illustration of the 
evolution of magnetic structure
at the time of flares in region C 
(group II, see Fig.~\ref{fig:0913X15event}d).
This figure is based on the data of the M6.1 flare on 12th as a 
representative of the flares in region C.
However, we believe that all the flares in region C 
occurred in a similar way.
(a) Simple sketch of magnetic structure in the active region
when the flare occurred. White and black indicate 
positive and negative polarities. 
The dashed line indicates the S-shaped neutral line and
the gray curve on the neutral line indicates the filament in 
region C.
(b) A cross section of magnetic structure along line AB 
shown in Fig. \ref{fig:regionC_structure}a.
Solid lines are magnetic field lines and 
hatched area indicates the filament. 
The negative element pointed by an arrow is considered to be 
the key of this event. Between the field line 
connected to the element
and the loops overlying the filament a current sheet 
forms (dotted line).
Note that the positive element on the left side of the pointed 
negative element was not on line AB in reality;
we consider that it was located in the positive side 
of the delta-type spot.
(c) In the current sheet, magnetic reconnection occurred 
and the loops overlying the filament 
brightened and were lengthened.
The bright loop is indicated by the thick curve.
The gray arrow indicates an ascending motion of the filament.
(d) As the reconnection proceeds, 
arcades with increasing height allow the filament to ascend.
}
\label{fig:regionC_structure}
\end{figure}

%region C flares
In region C, the negative elements 
flowed out from the negative part of
the delta-type spot during 12th and 13th.
When small flares occurred in region C, 
such as the M1.5 flare at 05:05 UT on Sep.~12th 
and the M6.1 flare at 09:03 UT on Sep.~12th,
the magnetic field lines connecting to these
negative elements reconnected to 
the loops overlying the filaments in region C;
a schematic illustration is shown in 
Figure \ref{fig:regionC_structure}.
Note that the positive element on the left side of the pointed 
negative element shown in Figures 
\ref{fig:regionC_structure}b, c, and d
was not on line AB shown in Figures 
\ref{fig:regionC_structure}a in reality;
we consider that it was located in the positive side 
of the delta-type spot.
During the events in region C,
bright loops crossing over the dark filament were observed.
It is the evidence that the loops overlying the filament
were involved in the reconnection process as shown in 
Figure \ref{fig:regionC_structure}.
As mentioned in \S \ref{section:regionC},
no eruption was observed during all but 
the M6.1 flare at 09:03 on Sep.~12th;
the M6.1 flare was associated with 
a small filament eruption in region C which
appeared to evolve into a CME. 
We consider that
most events only changed the equilibrium state of the filament
in region C, while
the perturbation at the time of the M6.1 flare was 
strong enough to trigger the eruption of the filament.
However, we consider that 
all these flares in region C (group II events)
occurred in essentially the same way as the mechanism shown in  
Figure \ref{fig:regionC_structure}.

% C & T (in the same way)
As for the flares in region T (group III events),
we interpret also in a similar way.
Although group II and group III were apparently independent events,
they were located along a single magnetic neutral line 
(the S-shaped neutral line; see Figure \ref{fig:0913X15event}f)
in the active region.
Moreover, during the X1.5 event,
throughout the east part of the S-shaped neutral line
flare ribbons were observed.
Therefore, we suppose that these filaments were
different parts of one flux rope or sheared field
along the S-shaped neutral line,
and similar mechanisms for small flares along the neutral line
(group II and III events) are suggested.

\begin{figure}[htpb]
\begin{tabular}{cccc}
\includegraphics[width=0.18\textwidth]{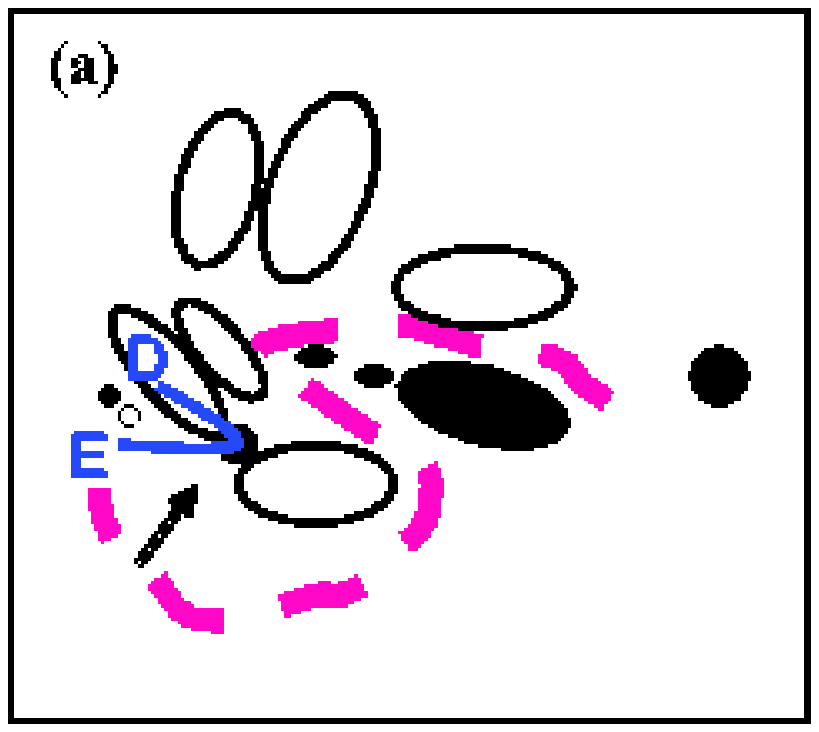} & %&&\\
\includegraphics[width=0.25\textwidth]{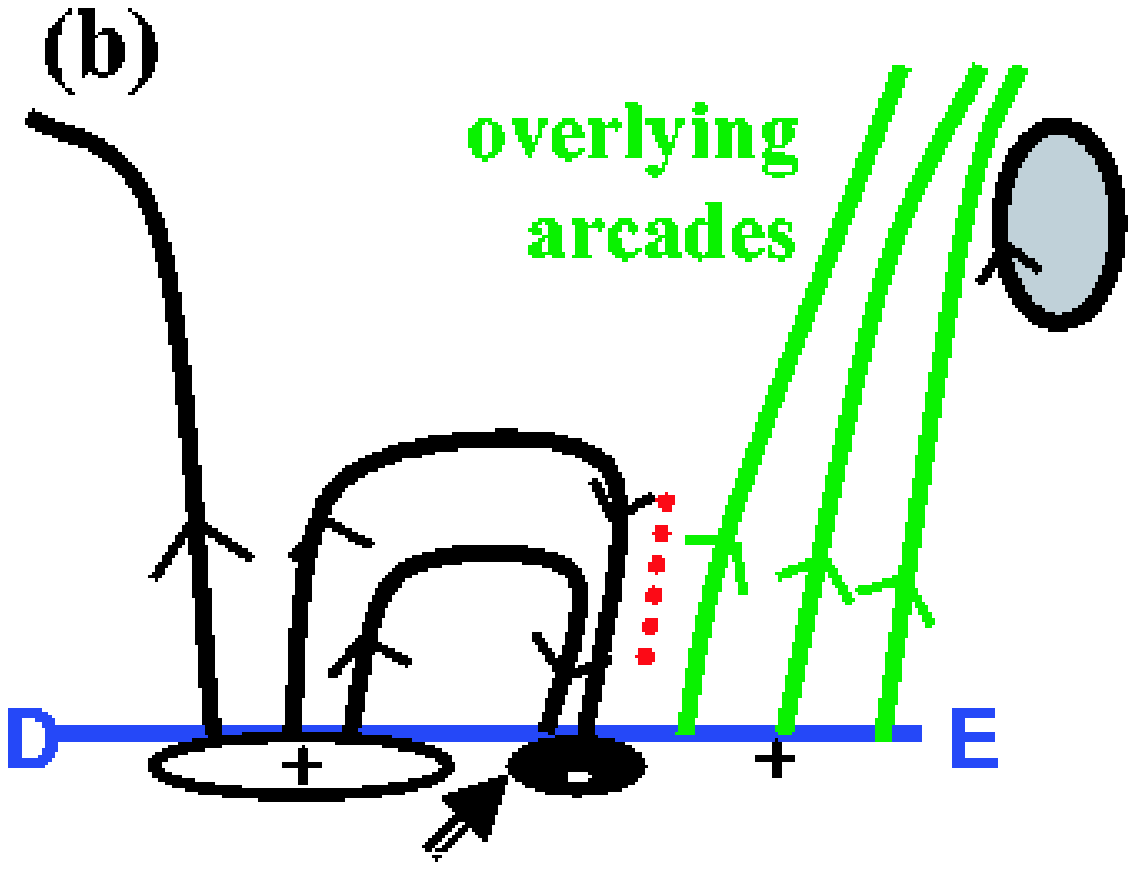} &
\includegraphics[width=0.25\textwidth]{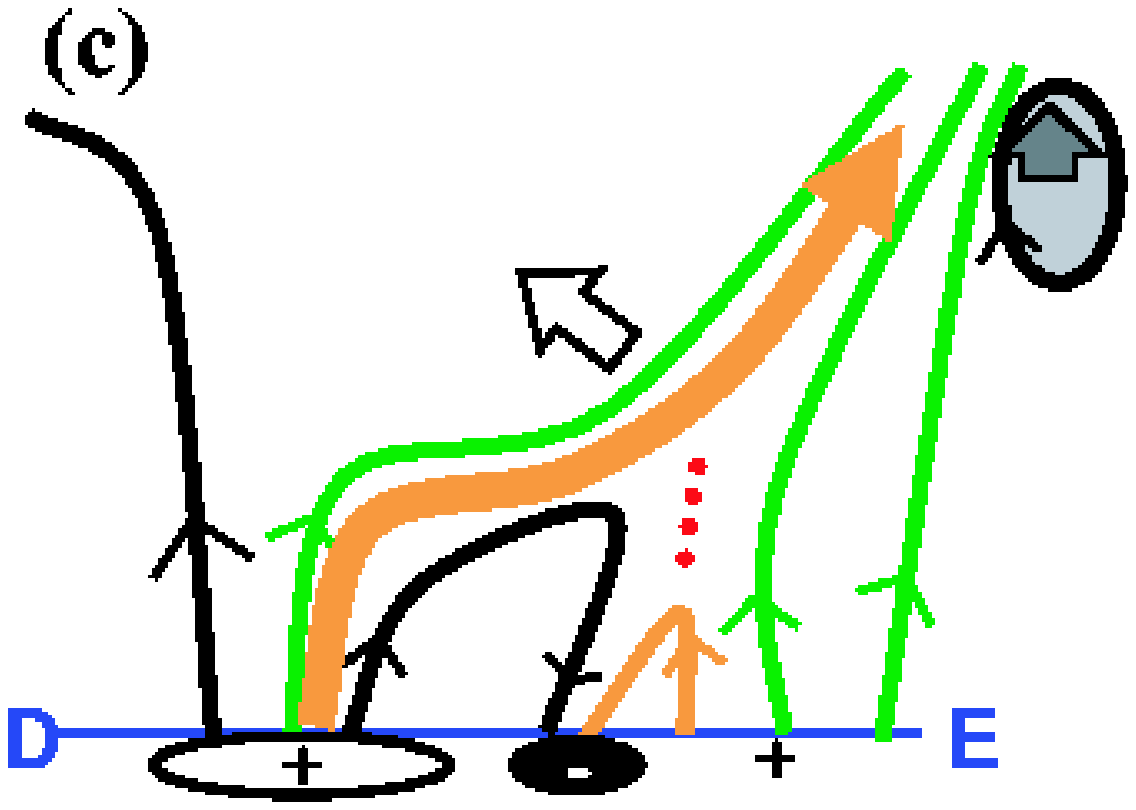} &
\includegraphics[width=0.25\textwidth]{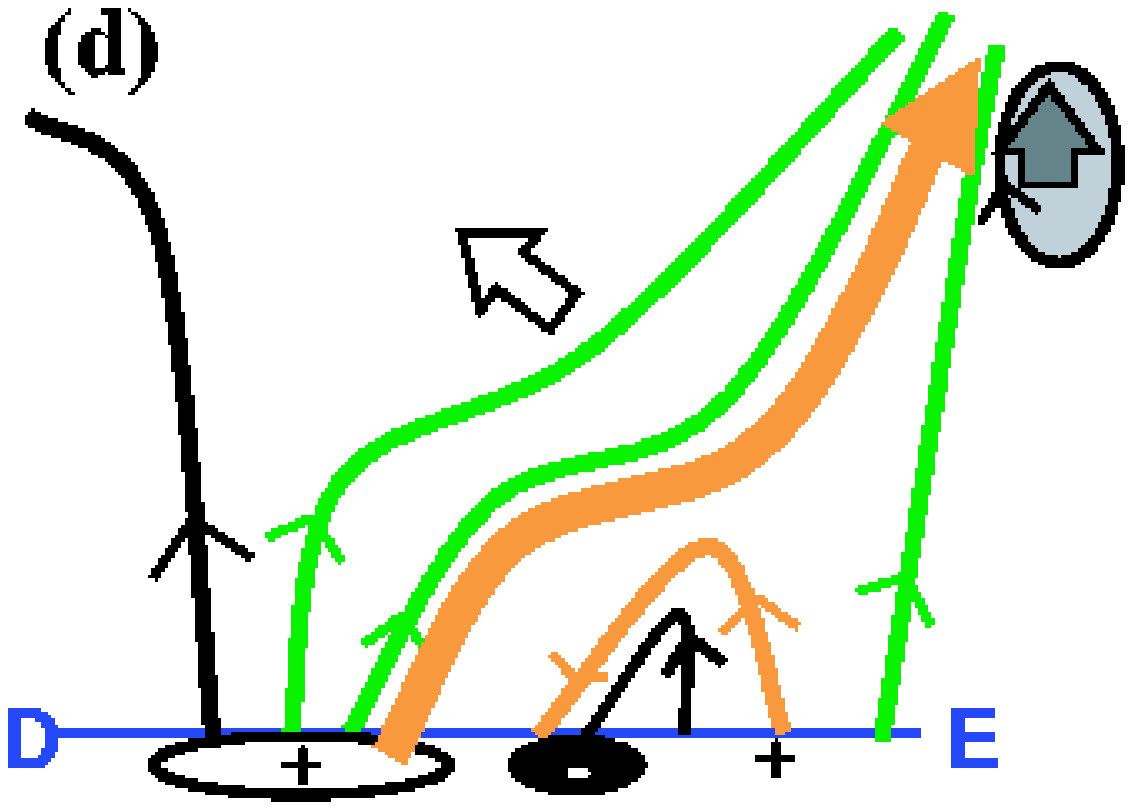} \\
&
\includegraphics[width=0.25\textwidth]{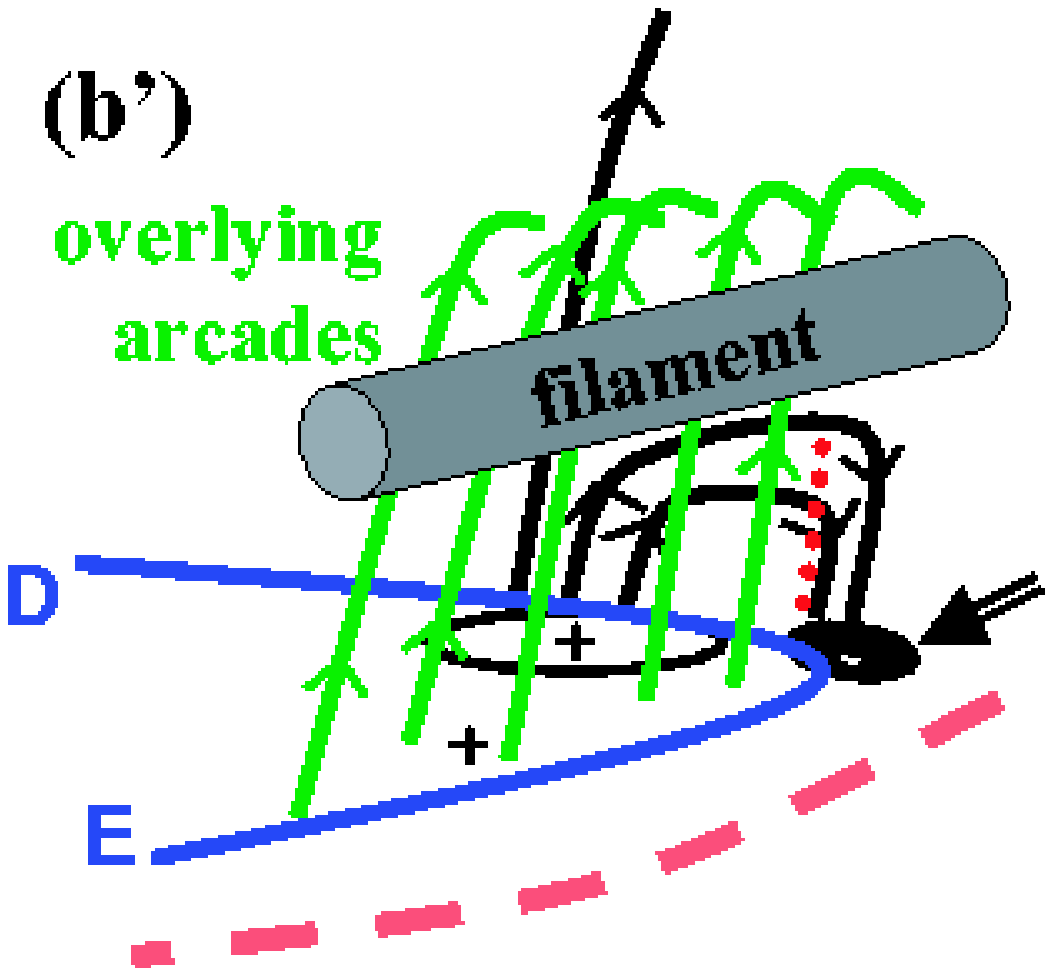} &
\includegraphics[width=0.25\textwidth]{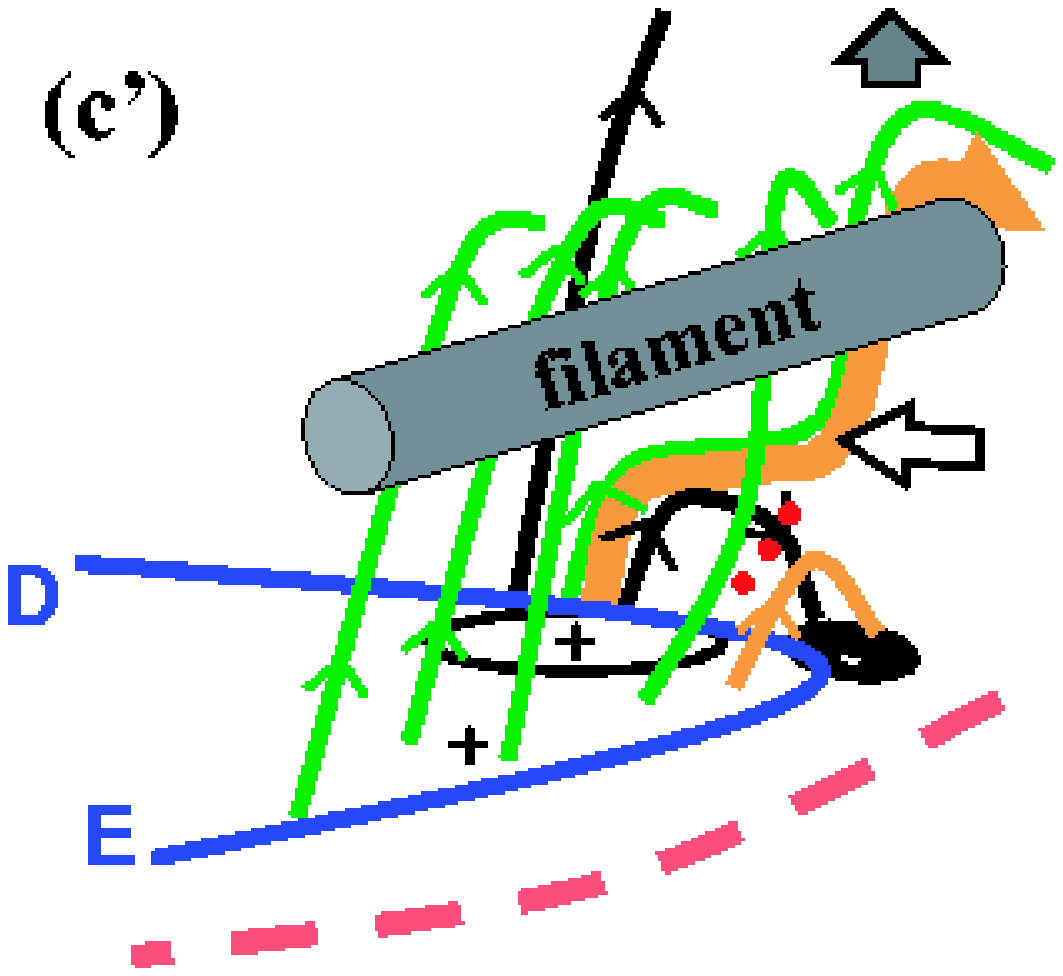} &
\includegraphics[width=0.25\textwidth]{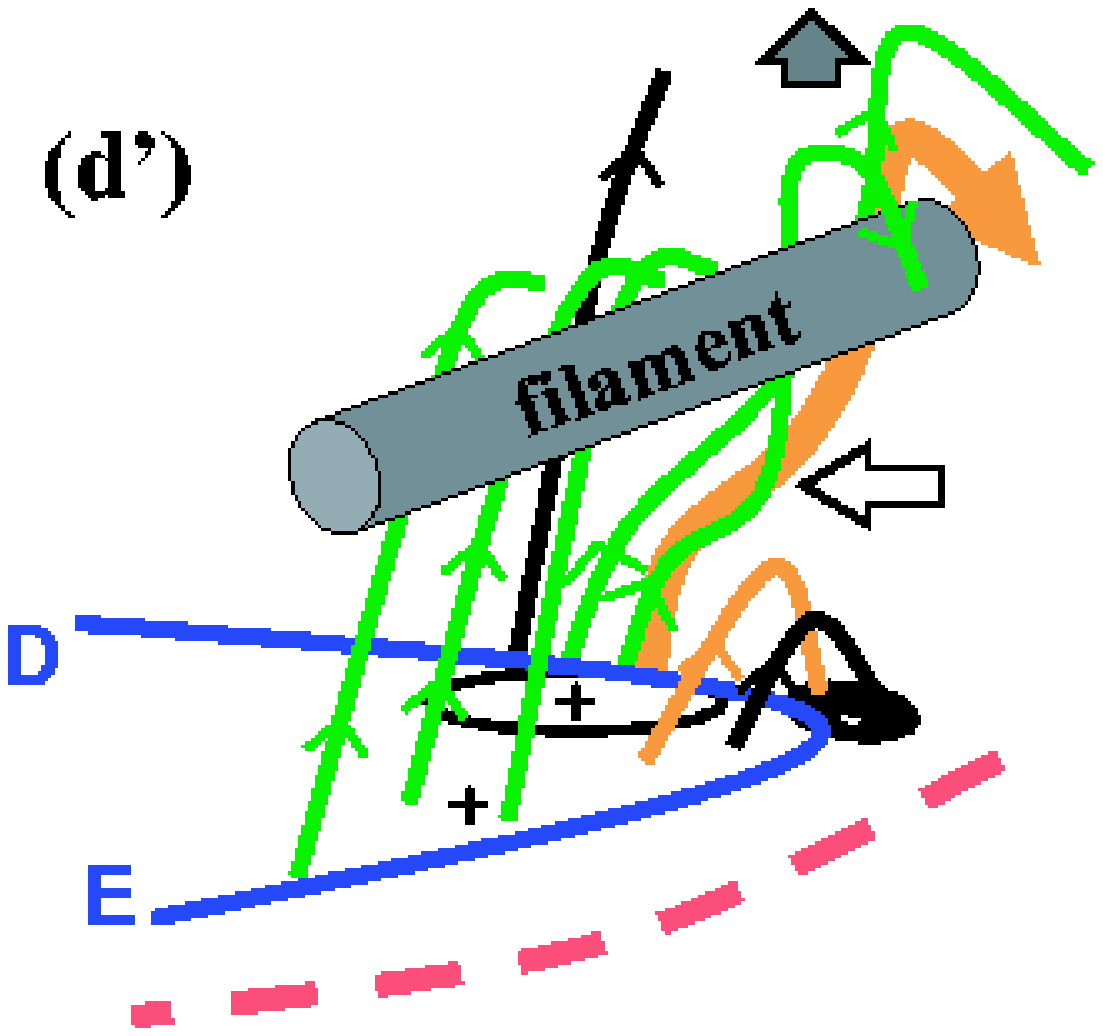} \\
\end{tabular}
\caption[Evolution of magnetic structure in region T. (the C2.9 flare)]
{Schematic illustration of the evolution of magnetic structure
at the time of flares in region T (group III, see Fig.~\ref{fig:0913X15event}d).
This figure is based on the data of the C2.9 flare on Sep.~13th 
as a representative of the flares in region T.
(a) Simple sketch of magnetic structure in the active region 
when the flare occurred.
White and black indicate positive and negative polarities.
The dashed line indicates the S-shaped neutral line.
(b) A cross section of magnetic structure along curve 
DE shown in Fig.~\ref{fig:regionT_structure}a.
In Figs.~\ref{fig:regionT_structure}b, c, and d, 
curve DE is unbent in a straight line.
The negative element pointed out by an arrow is 
considered to be the key to this event.
This element emerged into region T and moved toward southwest.
Hatched area indicates filament 2.
Between the overlying loops and the emerging flux 
a current sheet formed (red dotted line).
(c) In the current sheet, magnetic reconnection occurred.
A small bright loop (thin orange curve)
appeared as the left footpoint put on 
the negative element. The overlying arcade was lengthened and 
the filament anchored by the arcade ascended as indicated by 
a gray arrow.
A bright jet-like feature indicated by orange arrow
showed a whip-like motion as the loop stretched.
An outlined arrow indicates this whip-like motion.
(d) As the reconnection proceeded, jets with whip-like motion 
gradually moved eastward 
and the filament continued to move upward.
(b')Three-dimensional schematics of the magnetic structure 
along curve DE. Green lines indicate loops overlying 
filament 2 indicated by a gray cylinder.
(c')This panel shows the same phase as 
Fig.~\ref{fig:regionT_structure}c.
(d')This panel shows the same phase as 
Fig.~\ref{fig:regionT_structure}d.
}
\label{fig:regionT_structure}
\end{figure}

% slow-rise + C2.9
As the reconnection associated with these small flares
lengthened overlying loops, 
the filaments changed its equilibrium state and
appeared to deviate from the neutral line.
Finally, the filaments erupted after the C2.9 flare 
that occurred at 19:05 UT on Sep.~13th in region T.
We suppose that this small flare was a direct trigger of
the filament eruption (the X1.5 event). 
At that time the filaments were probably 
very close to the critical point for loss of equilibrium,
and were caused by the C2.9 flare to erupt.
Magnetic configuration of the C2.9 flare on Sep.~13th 
is shown in Figure \ref{fig:regionT_structure}.
The emerging flux whose footpoint located on the
flowing negative element interacted with the loops
overlying filament 2. The reconnected loop brightened 
and a jet-like feature crossing over the filament was observed.
This jet-like feature showed a whip-like motion
as the magnetic reconnection proceeded.

The characteristic features observed during Sep.~12th and 13th 
are summarized as follows:
(1) Filament 1 continuously ascended slowly.
(2) Lots of M- and C-class flares occurred at around the footpoints
of filaments; however, no eruption was observed during all but the 
M6.1 flare in region C on Sep.~12th.
(3) Although the C2.9 flare on 13th showed the 
similar magnetic structure to
the preceding M- and C-class events and was not 
a strong event in particular,
it seemed to trigger the catastrophic eruption (the X1.5 event).
We suggest that a series of 
small flares triggered the filament eruption
through the process of lengthening the overlying arcade
by magnetic reconnection.
This is consistent with the emerging flux  
triggering mechanism suggested by \citet{2000ApJ...545..524C} 
in that magnetic reconnection that occurred
in a filament channel triggers a filament eruption.
Note that \citet{1993SoPh..143..119W} also suggested
two-step magnetic reconnection:
The first step of reconnection is a slow reconnection in the lower
atmosphere that is observed as flux cancellation, while
the flare energy release comes directly from the second step
reconnection that is the fast reconnection higher in the corona.
However, 
the fact that lots of small flares occurred around the 
footpoints of filaments but no eruption was observed except 
the M6.1 flare on Sep.~12th (feature (2) described above) 
indicates that
magnetic reconnection at the footpoints of filaments is not a  
sufficient condition for the eruption.
In order to trigger the eruption, 
such reconnection must occur when the filament is
close to the critical point for loss of equilibrium; that is  
why the relatively small (C2.9)
flare could trigger the eruption.
Since in the initial condition of the 
simulation of \citet{2000ApJ...545..524C},  
the flux rope was already set
in an equilibrium state very close to instability or
loss of equilibrium, once magnetic flux emerges 
the flux rope can erupt immediately.
A loss of equilibrium model was proposed 
by \citet{1990JGR....9511919F} and \citet{1991ApJ...373..294F}.
\citet{1990JGR....9511919F} confirmed
by a numerical simulation 
that when the filament current exceeds a  
critical value, the stable
configuration containing the filament 
loses equilibrium and the filament may erupt.
We suggest the slow and long-lasting 
ascending motion of the filament
presented in this paper corresponds 
to the change of the equilibrium height of
the filament; the filament approaches to the critical point from  
initially stable
equilibrium, as described in the loss of equilibrium model.
For our X1.5 event, such change was due to 
the series of small flares involving the
overlying arcade of the filament.

\begin{figure}[htpb]
%\plottwo{./discussion/C29_structure_before_wb.eps}
%{./discussion/C29_structure_after_wb.eps}
\plottwo{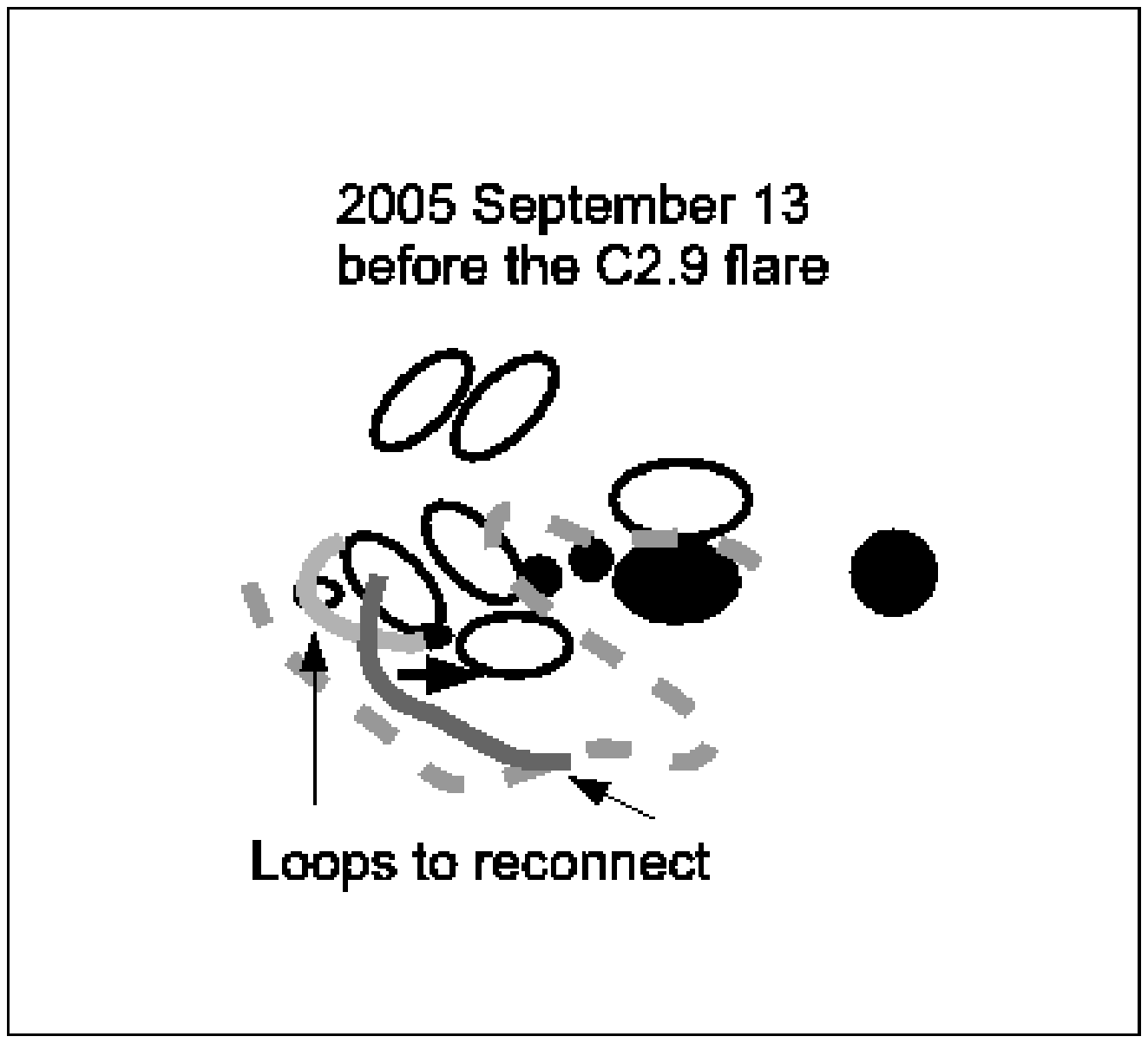}{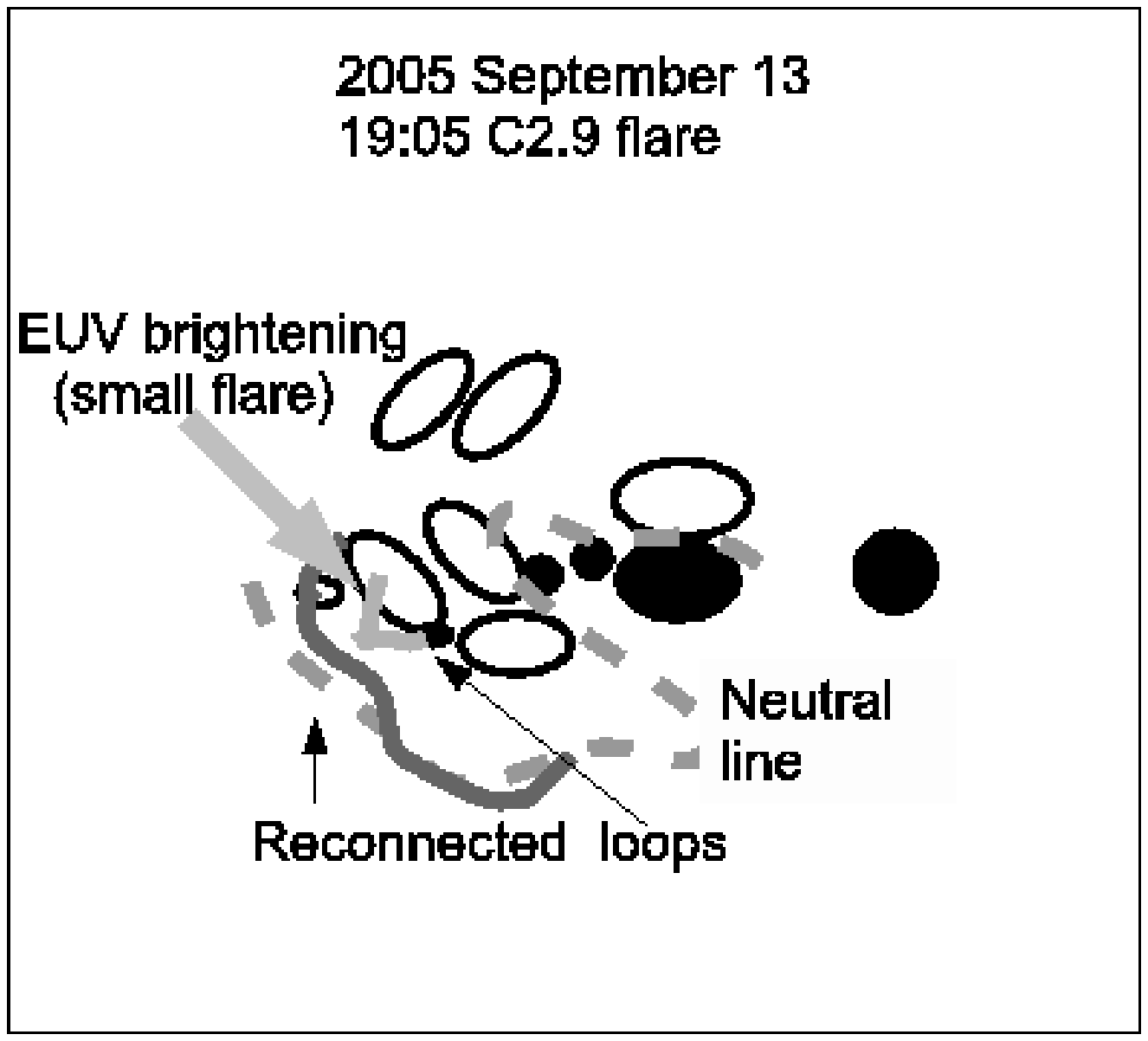}
\caption[Alternative view of the C2.9 flare.]
{Alternative interpretation of the C2.9 flare at 19:05 UT 
on September 13.
White and black indicate positive and negative polarities.
The dashed line indicates the S-shaped neutral line.
In the left panel, 
two magnetic lines crossed each other; 
this is the view before the small flare.
As the negative polarity elements moved westward,
the light gray magnetic line was pressed against 
the dark gray line
and finally they reconnected (see the right panel).
This dark gray line indicates filament 2 and
owing to the reconnection it became longer.
Therefore, filament 2 could not be held 
and it erupted with filament 1.}
\label{fig:C29_structure}
\end{figure}

It has to be mentioned that 
we can interpret these processes in another way.
In Figure \ref{fig:C29_structure}, 
we show the structure of the C2.9 flare that
occurred right before the X1.5 event.
In this scenario, we consider the moving magnetic elements played 
the key role in triggering the filament eruption;
at the footpoint of the flux rope labeled as filament 2
reconnection occurred and it made
the flux rope itself lengthen directly.
Then filament 2 erupted and filament 1 that was already
approaching certain critical point 
also erupted in a consequence of eruption of filament 2.

Another interesting feature to be noted here is 
a rising motion of bright structure during the slow rise of 
the filament indicated by an arrow in the time slice shown
in Figure \ref{fig:long_timeslice}.
This rising structure corresponds to an apparently swelling loop 
in {\it TRACE} image. This may be related to `bugle',
which was reported by \citet{1993JGR....9813177H}.
Bugles were observed as 
brightening and swelling of the bright belt of coronal 
streamers several days before a CME, and
they suddenly disappeared as the CMEs occurred.
They were named so because the shape looks like a bugle
on synoptic maps at a given height in the corona.
The timescale of bugles is the same order as 
the slow swelling motion in the corona found in our event,
although the size of bugle is much larger.
Perhaps the swelling motion is the lower coronal counterpart of a 
bugle.
One of the bugles reported by \citet{1993JGR....9813177H} 
was investigated by \citet{1995JGR...100.3355F}.
During the development of the bugle,
a new magnetic flux was observed to emerge below the 
large-scale arcades. 
As \citet{2000ApJ...545..524C} suggested,
an emerging flux can trigger filament eruptions
through the reconnection with the magnetic field lines 
around the filament.
In our event, we found systematic motions of magnetic elements 
below the filament. Such moving magnetic elements may or may not 
be related to newly emerging fluxes, but they also can induce the 
magnetic reconnection that changes the equilibrium state of the
filament and triggers its eruption.
Therefore, we suggest that (1) both emerging fluxes 
and moving magnetic elements in the vicinity of filament 
neutral line can lead a filament to 
approach the loss of equilibrium, and
(2) both the bugles in the high corona and the swelling loops in 
the low corona are the manifestations of the evolution of coronal
field that is approaching to the loss of equilibrium (eruption).

Using multi-wavelength data of these events,
such as {\it TRACE} EUV 195 \AA \ and 1600 \AA \  data,
MDI magnetograms, and BBSO and SMART H$\alpha$ data,
we tried to investigate extensively 
what happened in the active region
before the filament eruption.
Although we find several interesting features,
we cannot find definitive conclusion of what directly triggers
such flares in the active region,
since the magnetic structure of flares were extremely complex
in this region.
It is not clear whether 
one can know the equilibrium properties of a filament
(flux rope) from a ``snap shot" of the active region 
by, e.g., non-linear force free extrapolation of 
a photospheric magnetogram. Such approaches may work, 
but possibly one needs the evolutionary 
history of the active region
for an accurate diagnostics.
Comparison of coronal observations and the numerical
models of coronal field calculated from observed magnetograms have
been done extensively (e.g.,\citealt{1997SoPh..174...65Y}). 
We stress the
importance of studying the long-term (namely more than a few days)
temporal evolution of the photospheric and coronal magnetic field
that eventually produces an eruption.
The Solar Optical Telescope onboard Hinode satellite obtains
accurate vector magnetograms with uninterrupted observation, and
therefore will provide ideal data sets for this purpose.

%%%%%%%%%%%%%%%%%%%%%%%%%%%%%%%%%%%%%%
\section{Conclusion}
\label{sec:10808conclusion}
Active region NOAA 10808
appeared in September 2005 and exhibited extraordinary 
flare activity which was the most active in Solar Cycle 23.
Using EUV, H$\alpha$, and magnetogram data
taken before and during eruptive events accompanied 
by an X1.5 flare (the X1.5 event),
we investigate the processes leading to the catastrophic eruption.
We find several interesting features to 
understand the physical mechanism of the X1.5 event.
The main points are as follows:
\begin{itemize}
\item
The filament which erupted during the X1.5 event
slowly ascended over two days before its eruption.
The speed of the ascending motion is approximately
$0.1 {\rm km \ s^{-1}}$.

\item
While the filament was ascending slowly,
lots of M- and C-class flares were observed close to 
the footpoints of the filament.
At the sites where the flares frequently occurred,
magnetic elements emerged and moved distinctively.
\end{itemize}

On the basis of these observational facts,
we discuss the triggering mechanism of the
filament eruption on 2005 September 13.
We suggest that many small flares that occurred 
in the vicinity of the filament
played a role in %loosening the magnetic lines
changing the topology of the magnetic field lines 
overlying the filament through small scale reconnection. 
Over two days,
they changed its equilibrium state gradually 
and allowed the filament to ascend slowly.
In the end, a C2.9 flare that occurred just before 
the X1.5 event was considered to lead to 
the catastrophic filament eruption directly.

\acknowledgments
This work is supported by the Grant-in-Aid for 
the 21st Century COE ``Center for Diversity and 
Universality in Physics" from the Ministry of 
Education, Culture, Sports, Science and Technology (MEXT) of Japan,
and by the Grant-in-Aid for Creative 
Scientific Research ``The Basic Study of Space Weather Prediction''
(17GS0208, Head Investigator: K. Shibata) from MEXT.
K.N., H.I., and T.J.O. are supported by the Research Fellowship 
from the Japan Society for the Promotion of Science for 
Young Scientists.
%% Isobe & Tripathi 2006 A&A
{\itshape TRACE} is a NASA Small Explorer mission.
%%% Chifor et al. 2006
{\itshape SOHO} is a project operated by 
the European Space Agency and 
the US National Aeronautics and Space Administration.
%%% reconnection rate.
We thank the Big Bear Solar Observatory (BBSO) for providing access
to H$\alpha$ data.
Data analysis was carried out on the computer system
at the Nobeyama Solar Radio Observatory of the National 
Astronomical Observatory of Japan.

\bibliography{apj-jour,fil_erup,instrument}

\clearpage
%%% TABLE

\begin{deluxetable}{cc}
\tabletypesize{\scriptsize}
\tablecaption{Time-line summary of 
the 2005 September 13 X1.5 event\label{tab:050913X15}}
\tablecolumns{2}
\tablewidth{0pt}
\tablehead{
\colhead{Time(UT)} &
\colhead{Event} }
\startdata
18:40 & Brightening in region T begins. (195 \AA) \\
19:05 & {\it GOES} Soft X-Ray flux peak at C2.9\tablenotemark{a} \\
19:22 & Brightening in region C (H$\alpha-0.6$\AA)\\
19:24 & Filament 1 begins to rise (H$\alpha-0.6$ \AA)\\
19:24 & Filament 2 begins to brighten and rise (195 \AA)\\
19:27 & {\it GOES} Soft X-Ray flux first peak at X1.5\\
20:00 & {\it GOES} second peak, faint eruption (195 \AA) \\
20:00 & Halo CME with the speed of $1866 \rm{km \ s^{-1}}$ 
(LASCO C2)\\
\enddata
\tablenotetext{a}{We consider this event as a direct trigger of 
the filament eruption; see text.}
\end{deluxetable}

\clearpage
\begin{deluxetable}{ccccccc}
\tabletypesize{\scriptsize}
\tablecaption{X-, M-, and C-class flares on 
Sep.~12th and 13th \label{tab:MCclassflares}}
\tablecolumns{7}
\tablewidth{0pt}
\tablehead{
\colhead{} & 
\multicolumn{3}{c}{time\tablenotemark{a}} & \colhead{} &\colhead{} &\colhead{} \\
\cline{2-4}  \\
\colhead{date} &
\colhead{start} & 
\colhead{peak}    & 
\colhead{end}    & 
\colhead{GOES class} &
\colhead{CME} &
\colhead{location [group]}}
\startdata
Sep.~12th& 00:45 & 00:49 & 00:53  & C3.3 & None& region C 
[group II] \\ 
&  02:42 &02:48 &02:53 & C2.0  & None& region C  [group II]\\ 
&   04:49 &05:05& 05:27 & M1.5  &  None& region C  [group II]\\
&  06:56 &07:01& 07:05&  M1.3& None& region T (failed eruption) [group III]\\
&   08:37& 09:03& 09:20 & M6.1& CME & region C (eruption) [group II]\\  
&   15:33& 15:37& 15:42 & C1.1& None& region C [group II]\\
&   16:29& 16:33& 16:35&  C1.1& None& region T [group III]\\ 
&   16:35& 16:38& 16:40&  C1.2& None& region T [group III]\\
&   19:28& 19:40& 19:42&  C3.2& None&  jet  [group IV]\\
&   20:05& 20:09& 20:11&  M1.5& None& region T [group III]\\
&   22:07& 22:25& 22:42&  C7.2& None& region C [group II]\\
&   22:57& 23:01& 23:03&  C5.6& None& to the southwest of delta-type spot\\
&   23:12& 23:17& 23:23&  C5.5&  None& region T [group III]\\
Sep.~13th&  00:53& 00:58& 01:02&  C4.3&  None& region T [group III]\\
&    03:21& 03:27& 03:31&  C3.4& None& region T [group III]\\
&  03:57& 04:00& 04:02& C1.5 & None& jet [group IV]\\
&    04:15& 04:18& 04:20&  C1.9& None& region T [group III]\\
&    04:37& 04:45& 04:54&  C5.1& None&region T  [group III]\\
&    06:31& 06:35& 06:37&  C1.2& None& region T [group III]\\ 
&    08:24& 08:28& 08:31&  C1.8& None& region T? [group III?]\\
&    10:41& 11:21& 11:24&  M1.3& None& region C [group II]\\  
&  13:11& 13:43& 13:53&  C4.5& None& region C [group II]\\ 
&   18:49& 19:05& 19:15&  C2.9& None&region T [group III]\\ 
&   19:19& 19:27& 20:57&  X1.5& halo CME& 
filament eruption along NL [group I]\\ 
& 23:15& 23:22& 23:30&  X1.7\tablenotemark{b}& CME & above the delta-type spot\\
\enddata
\tablenotetext{a}{The timescale of flares are based on the data 
of {\it GOES} X-ray flux. 
See http://www.ngdc.noaa.gov/stp/SOLAR/ftpsolarflares.html}
\tablenotetext{b}{In this study, we do not refer to this event.}
\end{deluxetable}

\clearpage
%%%% FIGURE

%%section{overview of the AR}

%%%%%%%%%%%%%%%%%%%%%%%%%%%%%%%%%%%%%%
%%section{09.13 X1.5}
%%%%%%%%%%%%%%%%%%%%%%%%%%%%%%%%%%%%%%

%%section{long-time evolution of filament}

\clearpage

\clearpage

%%%%%%%%%%%%%%%%%%%%%%
%%section{discussion}
%%%%%%%%%%%%%%%%%%%%%%

\end{document}